\documentclass[journal=cmatex,manuscript=article,layout=traditional]{achemso}
\usepackage[version=3]{mhchem} % Formula subscripts using \ce{}
\usepackage[]{hyperref}
\usepackage[nameinlink,noabbrev]{cleveref}
\usepackage{siunitx}
\usepackage{xcolor}
\usepackage{multirow}

\newcommand{\mixedsystem}[2]{\ce{CsPb(I_{#1}Br_{#2})_{3}}}
\newcommand{\supercell}[3]{$#1\times#2\times#3$}

\author{Viren Tyagi}
    \affiliation{Department of Applied Physics and Science Education, Eindhoven University of Technology, 5600 MB, Eindhoven, The Netherlands}
\author{Mike Pols}
    \affiliation{Department of Applied Physics and Science Education, Eindhoven University of Technology, 5600 MB, Eindhoven, The Netherlands}
\author{Geert Brocks}
    \affiliation{Department of Applied Physics and Science Education, Eindhoven University of Technology, 5600 MB, Eindhoven, The Netherlands}
    \alsoaffiliation{Computational Chemical Physics, Faculty of Science and Technology and MESA+ Institute for Nanotechnology, University of Twente, 7500 AE, Enschede, The Netherlands}
\author{Shuxia Tao}
    \affiliation{Department of Applied Physics and Science Education, Eindhoven University of Technology, 5600 MB, Eindhoven, The Netherlands}
    \email{s.x.tao@tue.nl}

\title{Halide diffusion in mixed-halide perovskites and heterojunctions}

\begin{document}

%\date{\today}

% Abstract should be written in the present tense and impersonal style (i.e., avoid we), and be at most 200 words long
\begin{abstract}
Migration of halide defects guides ion transport in metal halide perovskites and controls the kinetics of halide mixing and phase separation. We study the diffusion of halide vacancies and interstitials in \mixedsystem{x}{1-x} and CsPbI$_3$/CsPbBr$_3$ heterojunctions by molecular dynamics simulations using neural network potentials trained on density functional theory calculations. We observe enhanced diffusion of both vacancies and interstitials in the mixed halide compounds compared to the single halide ones, as well as a difference in mobility between Br and I ions in the mixed compound. Diffusion across heterojunctions is governed by the interface structure, where a Br-rich interface blocks migration of vacancies in particular, but an I-rich interface is permeable.
\end{abstract}

% Text: Please use section headings and subheadings as specified below. For communications, all section headings apart from Experimental Section should be removed
% Please make the first reference to a display item bold: \textbf{Figure 1}
% Do not abbreviate Figure, Equation, etc.; display items are always singular, i.e., Figure 1 and 2.
% Equations are always singular, i.e., Equation 1 and 2, and should be inserted using the {equation} environment, not as graphics
% Please do not use footnotes in the text, additional information can be added to the Reference list.

\section{Introduction}
Metal halide perovskites have gained traction in optoelectronic applications such as photovoltaics and light-emitting diodes \cite{liang2023role, Lin2018perovskite, Liu2019low}. One of their strong points is chemical flexibility, where the perovskite ABX$_3$ structure allows for mixing different organic or inorganic cations (A), metal cations (B), and halide anions (X). Through compositional alloying, the physical properties such as band gaps and band offsets can be tuned, and at the same time this stabilizes the perovskite structure \cite{Yi2016entropic,saliba2016cesium,Jung2017influence}. For instance, mixing \ce{Br} into \ce{CsPbI3} lowers the temperature at which it transitions to the $\alpha$ (black) phase, and is less prone to degrade into the photo-inactive $\delta$ (yellow) phase \cite{Nasstrom2020dependence}, thereby improving the phase stability of this material. The currently most stable metal halide perovskites ABX$_3$ comprise a mix of X halide anions, as well as organic/inorganic A cations \cite{liang2022selective,kessels2025tailoring}.

Although these alloys are stable under dark conditions, continuous light exposure commonly leads to demixing of \ce{I} and \ce{Br} ions \cite{hoke_reversible_2015,Wang2019suppressed,Draguta2017rationalizing} and consequently the formation of low band gap I-rich domains and high band gap Br-rich domains \cite{tang_local_2018}. In previous work we presented a thermodynamic theory for light-induced halide segregation, based on the free energy gain of photocarriers assembled in domains with a halide composition that has a lower band gap than the starting mixed phase \cite{chen2021unified,chen2022light}. The growth of such domains is then stimulated by this thermodynamic driving force. While the origin of halide phase segregation is thermodynamic, the kinetics of this process are poorly understood, and a fundamental strategy for blocking or slowing down the kinetics is missing. 

Microscopically, the most likely candidate for mass transport in these materials is the migration of halide defects \cite{barker_defect-assisted_2017}. Halide perovskites are relatively soft materials and have a substantial concentration of defects \cite{Xue2021first,Xue2022intrinsic,Wang2018defects}. Halide point defects are generally considered the dominant mobile species \cite{Ball2016defects}. The relative importance of vacancies versus interstitials seems to depend somewhat on the particular operational conditions and techniques used in different experimental studies, and may also reflect the different conditions under which the materials are synthesized \cite{eames_ionic_2015,yunative2016,liiodine2016,game2017ions,schmidt2026quantification}.

From a modeling perspective, the most straightforward calculations make use of density functional theory (DFT) calculations on static structures along predicted migration paths to obtain diffusion barriers \cite{eames_ionic_2015,mosconi2016light,sarkar2024effects,Sarkar2024ion}. Such migration paths may however be difficult to find. In addition, they may also not be sufficiently representative, as such modeling relies on one or a small number of transition states being adequate for describing diffusion. Irrespective of that point, halide perovskites display different phases as a function of temperature, so DFT calculations on $T=0$ K structures may not describe diffusion at room or higher temperatures very well.  

Molecular dynamics (MD) is a technique to simulate mass transport at the atomistic level directly, which also takes into account finite-temperature effects. Until recently, MD simulations have yielded limited information on diffusion in halide perovskites, as empirical model force fields tend to be too crude \cite{lahnsteiner2018finite}, and ab initio MD simulations are computationally too expensive to reach the required time and length scales for ion migration. Machine learned force fields (MLFFs) have advanced the capabilities of MD simulations, by combining a near ab initio accuracy with a computational cost only an order of magnitude above empirical models \cite{behler2021four,Unke2021machine,Wu2023applications}. Moreover, by carefully selecting the ab initio reference data for training, MLFFs can describe the dynamics of a material at various electronic states \cite{mosquera-lois2025point}.

In recent work we utilized atomic descriptor-based machine learned force fields \cite{bartok2013onrepresenting,Jinnouchi2019phase,Jinnouchi2019onthefly,Jinnouchi2020descriptors} to study the migration of different charged states of iodide interstitials and vacancies in \ce{CsPbI3} \cite{tyagi_tracing_2025}. We found that negative iodide interstitials and positive iodide vacancies, the most stable charge states for their respective defect type, migrate at similar rates at room temperature. In contrast, oppositely charged defects, i.e. positive iodide interstitials and negative iodide vacancies, migrate at much slower rates. 

In the current work, we focus on the ion migration in mixed \mixedsystem{x}{1-x} compounds, where diffusion of point defects may involve both I and Br interstitials, as well as vacancies, and can depend on the local I versus Br environment \cite{sarkar2024effects}. We perform ns-long molecular dynamics (MD) simulations using MLFFs to quantify the migration behavior of halide defects in mixed \mixedsystem{x}{1-x}. In particular, we focus on halide diffusion through interfaces between \ce{CsPbI3} and \ce{CsPbBr3} layers and study the initial phases of halide mixing. To conclude, we examine the stability of a \ce{CsPbI3} nanodomain embedded in a \ce{CsPbBr3} matrix and vice versa with respect to halide mixing at the interfaces induced by defect migration \cite{sarkar2024effects}. 

%We find that the preferential migrating species is different for halide interstitial and halide vacancy, and halide phase mixing in domains is initiated by different halide point defects depending on the nature of the interface.

%Several studies have been carried out to understand this phenomenon. For instance, it is widely reported that primary ion migration channels in mixed halide perovskites are of halide vacancies \cite{brennan_light-induced_2018,ruth_vacancy-mediated_2018,yoon_shift_2017}. 

%However,  Phase segregation is highly dependent on the local topography, and is primarily observed at grain boundaries \cite{tang_local_2018}. Further, in nanocrystals (NC), it has been shown that \ce{I} ions migrate to the surface, leading to the formation of a single \ce{CsPbBr3} NC after segregation \cite{feng_complete_2024}. On top of that, \ce{I}/\ce{Br}-rich nanodomains can also form during the growth of NC due to lattice strain induced by size mismatch between \ce{I} and \ce{Br} ions \cite{liu_discrete_2023}. 

% What you had below would be suited in the introduction for a computational journal, but not for a general journal. I have merged it with the methods section and made it shorter. If you think it is useful to have a more in-depth discussion on the method, then it can be done in the supporting information

\section{Methods}
MLFFs based on Gaussian approximation potentials (GAPs)~\cite{Jinnouchi2019onthefly, Jinnouchi2019phase} with smooth overlap of atomic position (SOAP) descriptors~\cite{bartok2013onrepresenting, Jinnouchi2020descriptors} can accurately capture the diffusion of point defects, as we demonstrated in previous work on CsPbI$_3$ \cite{tyagi_tracing_2025}. Likewise, they are suited to describe defect migration in CsPbBr$_3$, as shown in SI note 1. However, they suffer from a poor scaling with the number of different chemical species, such as occurs in mixed \mixedsystem{x}{1-x} compounds. 

Although, message passing neural network (MPNN) potentials \cite{schutt2017schnet,Batzner2022equivariant} overcome this shortcoming, global message passing gives them a poor computational scaling with system size. The Allegro neural network architecture \cite{Musaelian2023learning} keeps the message passing local and is E(3)-equivariant (e3nn) \cite{Geiger2022e3nn}, which restrains the scaling with size and complexity of the system, see the SI note 2. As a result of this, an efficient MLFF for long-time scale MD simulations on large systems. Here, we train an Allegro model, as implemented in the NequIP package \cite{Batzner2022equivariant}, to generate a neural network potential (NNP) for halide interstitial and vacancy defects in \mixedsystem{0.5}{0.5}. 

The datasets for training and validating the network are generated using the on-the-fly module as implemented in VASP \cite{Kresse1996efficient,Jinnouchi2019phase}, where structures are sampled from \SI{100}{ps} long MD runs. Energies, forces, and stress tensors are obtained from density functional theory (DFT) calculations with the PBE+D3-BJ exchange-correlation functional \cite{perdew1996generalized,Grimme2011effect}, for a set of structures selected through Bayesian inference, to ensure sufficient variance in the local environments. The training and validation sets are generated using \supercell{2}{2}{2} cubic supercells of {mixed \mixedsystem{0.5}{0.5} with \ce{I} and \ce{Br} positioned randomly on halide sites}, and one halide point defect for a negatively charged halide interstitial ($\mathrm{I_{X}^{-}}$), or a positively charged halide vacancy ($\mathrm{V_{X}^{+}}$). Details of the training procedure are explained in the SI note 3. 

\begin{figure}[!htbp]
    \includegraphics{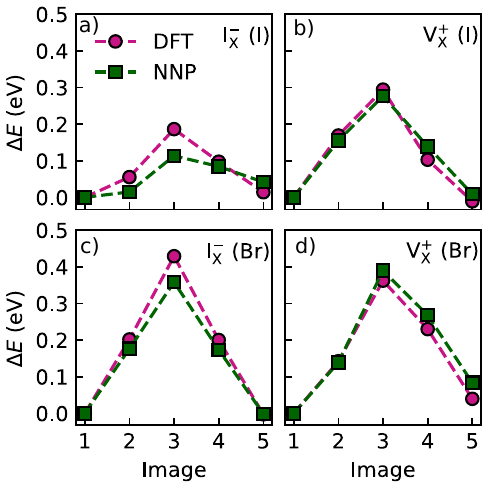}
    \caption{Energies along CI-NEB migration paths for I-mediated interstitial (a) and vacancy (b), and for Br-mediated interstitial (c) and vacancy (d) in \mixedsystem{0.5}{0.5} from DFT and NNP calculations. Zero is set at the energy minima.
    }
    \label{fig:NNP-NEB-energies}
\end{figure}

The accuracy of the NNP energies can be explored, for instance, by comparing migration barriers along paths generated using the climbing image nudged elastic band (CI-NEB) technique \cite{Henkelman2020climbing} with those calculated by DFT along the same paths. The computed energies for vacancies and interstitials in {mixed \mixedsystem{0.5}{0.5} systems with \ce{I} and \ce{Br} positioned randomly on halide sites} are given in Figure~\ref{fig:NNP-NEB-energies}a-d. The difference between the migration barriers calculated using DFT and the NNP is less than \SI{0.08}{eV} in all cases, indicating the NNP accurately probes the energy landscape for defect migration. The details of the CI-NEB calculations and calculated migration barriers are given in the SI note 4. 

The NNP model is used to perform ns-long MD simulations in LAMMPS \cite{thompson2022lammps} on \supercell{6}{6}{6} cubic supercells of bulk \mixedsystem{x}{1-x} with one halide point defect, interstitial or vacancy. To test the accuracy of the forces generated by the NNP, we sample structures from MD simulations at \SI{600}{K}, and calculate the forces by DFT. Comparing these to the NNP forces for the system for which the NNP was developed, i.e., {mixed \mixedsystem{0.5}{0.5} with \ce{I} and \ce{Br} positioned randomly on halide sites}, gives $R^{\mathrm{2}}=0.95$ and a mean absolute error (MAE) $\leq51$ \SI{}{meV/\angstrom} for forces acting on all atoms, and $R^{\mathrm{2}}\geq0.94$ and MAE $\leq56$ \SI{}{meV/\angstrom} for forces acting on atoms in the defect environment, see the SI note 4. Using the same NNP in MD simulations on the pure compounds, \ce{CsPbI3} or \ce{CsPbBr3}, gives a very similar accuracy in the forces, see SI figure S11, which is a strong indication of the transferability of the NNP among \mixedsystem{x}{1-x} compounds for $0\leq x\leq 1$.   

%Forces are compared for multiple structures sampled from MD simulations at \SI{600}{K} performed \supercell{6}{6}{6} cubic supercells with one halide point defect for pure \ce{CsPbI3}, randomly mixed \mixedsystem{0.5}{0.5}, and pure \ce{CsPbBr3}. For these comparisons, the $R^{\mathrm{2}}\geq0.95$ and the mean absolute error (MAE) $\leq51.55$ \SI{}{meV/\angstrom} for forces acting on all atoms, and the $R^{\mathrm{2}}\geq0.94$ and MAE $\leq55.69$ \SI{}{meV/\angstrom} for forces acting on atoms in the defect environment, for all systems, indicating the high accuracy and transferability of the NNP in calculating forces acting on the atoms for different halide defects and varying halide ratios. The details of these MD runs and the comparison of forces for NNP with DFT are given in SI note 3. 

%Along with energies and forces, we also test the transferability of this NNP by comparing the values of the diffusion coefficients of halide interstitial and halide vacancy at \SI{600}{K} ($D_{\mathrm{600\,K}}$), calculated from MD runs for pure systems, \ce{CsPbI3} and \ce{CsPbBr3}, with those obtained from MD runs using GAP-SOAP potentials (SI note 1) \cite{tyagi_tracing_2025}. The comparisons, see Table ??, illustrate the high transferability of the NNP in calculating diffusion coefficients for different halides. The details of these MD runs are given in SI note 3.

Besides studying the diffusion of halide defects in \mixedsystem{x}{1-x}, the NNP force field is also used to investigate migration of halide defects across interfaces between mono-halide layers, as well as halide mixing starting from mono-halide cubic nanodomains. The supercell size, temperature range, number of runs, and simulation time per run for all these systems are given in Table~\ref{tab:system-discription}. The volume is kept constant during these runs, where the lattice constants for different halide concentrations in  \mixedsystem{x}{1-x} are extracted from MD runs at constant temperature and pressure, see the SI note 5.

\begin{table*}
    \centering
    \begin{tabular}{|c c c c c|}
     \hline
     Model & Supercell & Temperature (\SI{}{K}) & time (\SI{}{ns}) & number of runs \\
     \hline
     Mixed halide & \supercell{6}{6}{6} & $\mathrm{500 - 600}$ & 2 & 5 \\
     Interfaces & \supercell{16}{6}{6} & $\mathrm{600}$ & 2 & 5 \\
     Nanodomains & \supercell{8}{8}{8} & $\mathrm{600}$ & 10 & 1 \\
     \hline
    \end{tabular}
    \caption{\textmd{The cubic supercell size, temperature range, simulation time per MD run, and number of runs used for the different systems in this study.}}
    \label{tab:system-discription}
\end{table*}

\section{Results}

% \begin{figure}
%     \includegraphics{2D-illustration-all-models.pdf}
%     \caption{2D illustration of a) randomly mixed \supercell{6}{6}{6} supercell of \mixedsystem{0.5}{0.5}, b) \supercell{16}{6}{6} supercell of \mixedsystem{x}{1-x} with \ce{I} and \ce{Br} phases segregated by 2 \ce{Br}-rich layers, and c) \supercell{8}{8}{8} supercell of \mixedsystem{x}{1-x} with a cubic \ce{Br} domain in \ce{I} lattice.}
%     \label{fig:2D-illustration-all-models}
% \end{figure}

\subsection{Diffusion in \mixedsystem{0.5}{0.5}}
Migration behavior is studied in {mixed \mixedsystem{0.5}{0.5} with \ce{I} and \ce{Br} positioned randomly on halide sites} with one halide point defect, vacancy or interstitial, added. We choose vacancy $\mathrm{V^{+}_\mathrm{X}}$ or interstitial $\mathrm{I^{-}}_\mathrm{X}$, X = I or Br, in their most stable charge states, as they occur under intrinsic or moderately doped conditions \cite{Xue2022intrinsic}. To ensure sufficient migration events for statistics, MD runs are performed at elevated temperatures in the range 500-600 K. The details of these production runs are given in the SI note 6. To compare with the pure systems, we also perform MD simulations for the pure systems \ce{CsPbI3} and \ce{CsPbBr3}. For the data obtained for \ce{CsPbI3}, see Ref.~\cite{tyagi_tracing_2025}. The training and production run details for \ce{CsPbBr3} and the diffusion behavior of other charge states of halide point defects are reported in SI note 1. 

In characterizing halide defect diffusion in these systems it makes no sense to trace a specific type of halide ion, as migration typically proceeds via kick-out processes. Therefore, we use both halide species to extract the diffusion coefficient using the Einstein relation, which links the slope of the mean square displacement to time
\begin{equation}
    D = \frac{1}{6}\lim_{t\rightarrow\infty}\frac{d}{dt} \Bigl \langle \sum_{i=1}^{N} \left\Vert \mathbf{r}_i(t+t_0)-\mathbf{r}_i(t_0)\right\Vert^{2}\Bigr \rangle_{t_{0}};\;\;\;t>t_0,
    \label{eqn:equation_MSD}
\end{equation}
where $N$ is the number of atoms of a particular atomic species in the simulation box, and $\mathbf{r}_i$ are the atomic positions. 

In the following we will label diffusion coefficients by interstitial or vacancy, according to the starting situation of each MD run. In the temperature and time ranges studied one observes multiple sequential kick-out events, so the actual species that is migrating can change over the trajectory.  However, when starting with a single vacancy (interstitial), then one vacancy (interstitial) remains present; no additional long-lived defects are created.

\begin{figure}[htb]
    \includegraphics{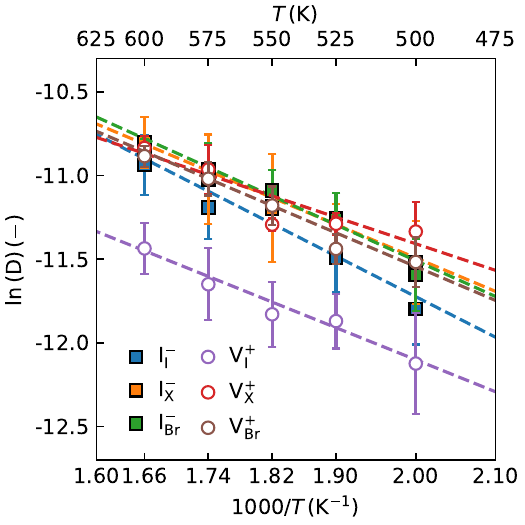}
    \caption{Temperature-dependent diffusion coefficients of halide interstitials ($\mathrm{I^{-}}$) and vacancies ($\mathrm{V^{+}}$) in pure \ce{CsPbI3} (\ce{I}), {mixed \mixedsystem{0.5}{0.5} with \ce{I} and \ce{Br} positioned randomly on halide sites} (\ce{X}), and pure \ce{CsPbBr3} (\ce{Br}). The filled square and open circle symbols represent halide interstitials and vacancies, respectively. The dashed lines represent fits to an Arrhenius expression, and the error bars represent the standard error in mean at each point.}
    \label{fig:diffusion-curve-all}
\end{figure}

Diffusion coefficients are given in Figure~\ref{fig:diffusion-curve-all}. They are extracted by applying Eq. \ref{eqn:equation_MSD} to the runs for the mixed-halide compound and similar systems created for the single-halide compounds. Comparing \ce{CsPbI3} and \ce{CsPbBr3} one observes that diffusion coefficients of Br and I interstitial, $\mathrm{I^{-}_{Br}}$ and $\mathrm{I^{-}_{I}}$, in the temperature range 500-600 K are very similar, i.e., within 30\% or so, with that of the I interstitial higher. The difference between the diffusion coefficients of Br and I vacancies, $\mathrm{V^{+}_{Br}}$ and $\mathrm{V^{+}_{I}}$, is slightly larger, approximately a factor of two, with that of the I vacancy being the larger of the two. In the mixed \mixedsystem{0.5}{0.5} system the two halide species can change positions dynamically. As a first analysis we look at the migration of the halide species X without distinguishing between Br and I. The diffusion coefficients for X interstitials and vacancies are remarkably close to those found in the pure systems, but slightly closer to the values found in \ce{CsPbBr3}, in particular for the vacancy. 

The diffusion coefficients can be fitted with an Arrhenius expression
\begin{equation}
    D = D_{0}\exp\left(-\frac{E_\mathrm{a}}{k_\mathrm{B}T}\right).
    \label{eqn:equationAR}
\end{equation}
The fits are given in Figure~\ref{fig:diffusion-curve-all}, and the extracted activation energies ($E_{\mathrm{a}}$) and pre-exponential factors ($D_{\mathrm{0}}$) are given in Table~\ref{tab:diffusion-data-all}. The error bars reflect the quality of the linear regression fit. 

{The activation energies are generally quite low and close to one another across all compositions, lying between \SI{0.13}{eV} and \SI{0.21}{eV}. While \ce{I} ions are larger than the \ce{Br} ions, the structures of CsPbI$_3$ and CsPbBr$_3$ scale accordingly, making the diffusion of I interstitials in CsPbI$_3$ similar to that of Br interstitials in CsPbBr$_3$, and the same holds for halide vacancy diffusion in both materials. The barriers for interstitial diffusion are somewhat higher than those for vacancy diffusion, but they are actually remarkably close. The diffusion barriers in the mixed \mixedsystem{0.5}{0.5} halide are somewhat lower than in the pure compounds. In addition, there is a slight difference in the migration behavior of I and Br defects in the mixed halide. We will come to this point below.} 

Extrapolating the Arrhenius expression to room temperature, $T=300$ K, one obtains values for the diffusion coefficients in the range $10^{-7}$-$10^{-6}$ cm$^2$s$^{-1}$, see Table~\ref{tab:diffusion-data-all}. The extrapolated diffusion coefficients of vacancies and interstitials in the mixed halide are approximately a factor of two larger than those in the pure compounds. In a similar fashion, the diffusion coefficients of vacancies are approximately a factor of two larger than those of interstitials.

\begin{table*}[htb]
    \centering
    \begin{tabular}{|c c c c c|}
        \hline
         {System} & Defect & ${E_\mathrm{a}\,\,\mathrm{(eV)}}$ & ${D_{0}\,\,(\times10^{-3}\,\mathrm{cm}^{2}\mathrm{s}^{-1})}$ & ${D_{300K}\,\,(\times10^{-7}\,\mathrm{cm}^{2}\mathrm{s}^{-1})}$\\
        \hline
         \multirow{2}{6em}{{\ce{CsPbI_{3}}}} & $\mathrm{I_{I}^{-}}$ & {$\mathrm{0.21\pm0.03}$} & {$\mathrm{1.08\pm0.77}$} & {$\mathrm{3.2}$}\\
          & $\mathrm{V_{I}^{+}}$ & {$\mathrm{0.16\pm0.03}$} & {$\mathrm{0.24\pm0.19}$} & {$\mathrm{4.9}$}\\
          \hline
          \multirow{2}{6em}{{\mixedsystem{0.5}{0.5}}} & 
         $\mathrm{I_{X}^{-}}$ & {$\mathrm{0.17\pm0.04}$} & {$\mathrm{0.54\pm0.46}$} & {$\mathrm{7.5}$}\\
         & $\mathrm{V_{X}^{+}}$ & {$\mathrm{0.13\pm0.02}$} & {$\mathrm{0.25\pm0.12}$} & {$\mathrm{16.4}$}\\
         \hline
         \multirow{2}{6em}{{\ce{CsPbBr_{3}}}} &
         $\mathrm{I_{Br}^{-}}$& {$\mathrm{0.19\pm0.03}$} & {$\mathrm{0.75\pm0.43}$} & {$\mathrm{4.8}$}\\
         & $\mathrm{V_{Br}^{+}}$ & {$\mathrm{0.17\pm0.02}$} & {$\mathrm{0.53\pm0.22}$} & {$\mathrm{7.4}$}\\ 
         \hline
    \end{tabular}
    \caption{\textmd{Activation energies (${E_\mathrm{a}}$) and pre-exponential factors (${D_{0}}$) extracted from the Arrhenius fits, and extrapolated diffusion constants (${D_{300\mathrm{K}}}$) at room temperature for halide interstitial ($\mathrm{I^{-}}$) and vacancy ($\mathrm{V^{+}}$) defects in pure \ce{CsPbI3} (\ce{I}), {mixed \mixedsystem{0.5}{0.5} with \ce{I} and \ce{Br} positioned randomly on halide sites} (\ce{X}), and pure \ce{CsPbBr3} (\ce{Br}).}}
    \label{tab:diffusion-data-all}
\end{table*}

%At room temperature, halide interstitials migrate slightly slower in the mixed system ($\mathrm{I_{X}^{-}}$) in pure \ce{CsPbI3} ($\mathrm{I_{I}^{-}}$) ($3.20\,\mathrm{vs}\,1.60\,\times10^{-7}$ \SI{}{cm^{2}s^{-1}}), and it is the fastest in pure \ce{CsPbBr3} ($\mathrm{I_{Br}^{-}}$) ($7.09\,\times10^{-7}$ \SI{}{cm^{2}s^{-1}}). Halide vacancies, however, migrate the fastest in the mixed system ($\mathrm{V_{X}^{+}}$) ($15.71\,\times10^{-7}$ \SI{}{cm^{2}s^{-1}}), and migrate slightly faster in pure \ce{CsPbBr3} ($\mathrm{V_{Br}^{+}}$) than in pure \ce{CsPbI3} ($\mathrm{V_{I}^{+}}$) ($\mathrm{I_{I}^{-}}$) ($7.38\,\mathrm{vs}\,4.92\,\times10^{-7}$ \SI{}{cm^{2}s^{-1}}). Finally, we also note that halide vacancies migrate about an order of magnitude faster than halide interstitials in the mixed systems ($15.71\,\mathrm{vs}\,1.60\,\times10^{-7}$ \SI{}{cm^{2}s^{-1}}).

\begin{figure}[htb]
    \includegraphics{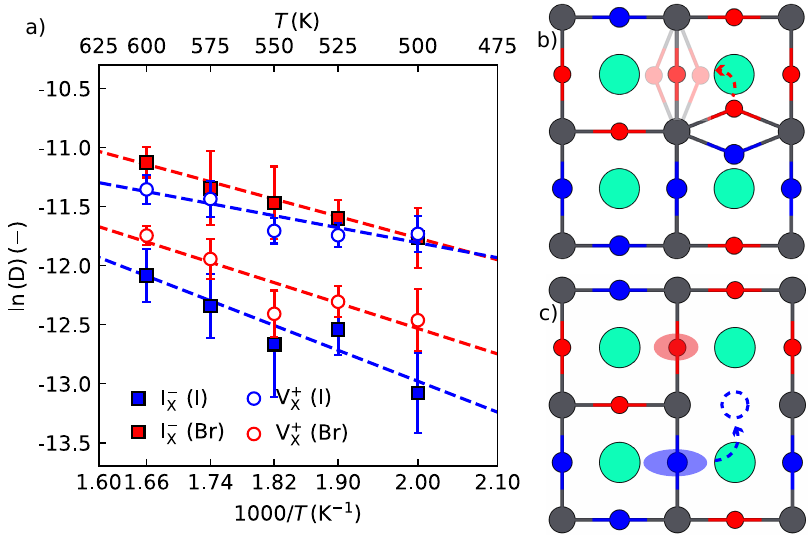}
    \caption{a) Halide species decomposed temperature-dependent diffusion coefficients of halide interstitials and vacancies in \mixedsystem{0.5}{0.5}. Blue is for \ce{I}, and red is for \ce{Br}. The filled square symbols represent halide interstitials, and the open circle symbols represent halide vacancies. The dashed lines represent the fits to an Arrhenius expression, and the error bars represent the standard error in mean at each point. Schematic representation of the diffusion paths for (b) halide interstitial and (c) halide vacancy defects in \mixedsystem{0.5}{0.5}. The arrows represent the migration directions of the ions.}
    \label{fig:diffusion-curve-migration-path-CsPbIBr}
\end{figure}

In principle, Eq. \ref{eqn:equation_MSD} also allows for decomposing the migration behavior of halide interstitials and vacancies in the \mixedsystem{x}{1-x} down to halide species, i.e., distinguishing between Br and I. Diffusion coefficients for each species are given in Figure~\ref{fig:diffusion-curve-migration-path-CsPbIBr}a). As it turns out, diffusion does not depend on the species, Br or I, initially chosen to represent interstitial of vacancy. As diffusion proceeds via kick-out processes, after a few of those events, the dominant species takes over. Regarding interstitials the Br species migrates faster than the I species, irrespective of the starting species of interstitial. Remarkably, for vacancies it is the other way around, i.e., I vacancies migrate faster than Br vacancies. 

Interstitials and vacancies have opposite charge, so in an electric field they move in opposite directions. As the negatively charged interstitials will be driven to the anode, it will lead to a surplus of Br there. At the same time, the positively charged vacancies migrate to the cathode, which promotes a net flux of I in the direction of the anode. In an electric field the two processes, interstitial- and vacancy-mediated diffusion, thus tend to compensate one another as a driving force for halide separation.  

Arrhenius fits, Eq. \ref{eqn:equationAR}, are shown in Figure~\ref{fig:diffusion-curve-migration-path-CsPbIBr}a, with the parameters and the extrapolation to room temperature given in Table~\ref{tab:diffusion-data-CsPbIBr}. These results confirm the trends discussed in the previous paragraph. Halide interstitials migrate an order of magnitude faster through \ce{Br} than through \ce{I}. Meanwhile, halide vacancies migrate faster through \ce{I} than through \ce{Br}. Focusing on the dominant migration mechanisms, migration of Br interstitials and I vacancies is roughly equally fast.

\begin{table*}
    \centering
    \begin{tabular}{|c c c c|}
     \hline
     System & ${E_\mathrm{a}\,\,\mathrm{(eV)}}$ & ${D_{0}\,\,(\times10^{-3}\,\mathrm{cm}^{2}\mathrm{s}^{-1})}$ & ${D_{300K}\,\,(\times10^{-7}\,\mathrm{cm}^{2}\mathrm{s}^{-1})}$\\
     \hline
     $\mathrm{I_{X}^{-}}$ (I) & {$\mathrm{0.22\pm0.06}$} & {$\mathrm{0.43\pm0.51}$} & {$\mathrm{0.87}$}\\
     $\mathrm{I_{X}^{-}}$ (Br) & {$\mathrm{0.15\pm0.04}$} & {$\mathrm{0.29\pm0.26}$} & {$\mathrm{8.8}$}\\
     $\mathrm{V_{X}^{+}}$ (I) & {$\mathrm{0.11\pm0.02}$} & {$\mathrm{0.08\pm0.05}$} & {$\mathrm{11.4}$}\\
     $\mathrm{V_{X}^{+}}$ (Br) & {$\mathrm{0.18\pm0.04}$} & {$\mathrm{0.24\pm0.19}$} & {$\mathrm{2.3}$}\\
     \hline
    \end{tabular}
    \caption{\textmd{Activation energies (${E_\mathrm{a}}$) and pre-exponential factors (${D_{0}}$) extracted from the Arrhenius fits to diffusion coefficients decomposed according to halide species in \mixedsystem{0.5}{0.5} and diffusion constants (${D_{300\mathrm{K}}}$) extrapolated to room temperature.}}
    \label{tab:diffusion-data-CsPbIBr}
\end{table*}

To rationalize halide diffusion in mixed \mixedsystem{0.5}{0.5}, we examine the defect geometries along the migration paths. The most stable configuration of a halide interstitial $\mathrm{I_X^-}$ is the bridge structure shown schematically in Figure~\ref{fig:diffusion-curve-migration-path-CsPbIBr}b) \cite{Meggiolaro2018first,Xue2021first}. In \mixedsystem{0.5}{0.5}, a \ce{Pb-BrBr-Pb} bridge configuration is more frequently observed than a \ce{Pb-IBr-Pb} bridge configuration, both of which are observed significantly more than \ce{Pb-II-Pb} bridge configuration, see the SI note 7. We conclude that the more prevalent a configuration is, the more stable it is. On the basis of occupation numbers, SI Figure S16, and assuming that these are determined by Boltzmann factors, we estimate an energy difference between the \ce{Pb-IBr-Pb} bridge and the \ce{Pb-BrBr-Pb} bridge of \SI{34.5}{meV}, and between the \ce{Pb-II-Pb} bridge and the \ce{Pb-BrBr-Pb} bridge of \SI{143.9}{meV} at \SI{600}{K}. As \ce{I} is much less likely to be found in an interstitial position than \ce{Br}, the latter is the main species that causes diffusion of interstitial defects, as also shown in Figure~\ref{fig:diffusion-curve-migration-path-CsPbIBr}. 

By contrast, the halide vacancy has a preference for diffusion by \ce{I}. The \ce{Pb-Br} bond is stronger than a \ce{Pb-I} bond, suggesting that the latter can be more easily broken to let an \ce{I} fill the vacancy, Figure~\ref{fig:diffusion-curve-migration-path-CsPbIBr}c). Consistent with that notion, the \ce{I} ions fluctuate more around their mean position, indicating a flatter energy landscape, than \ce{Br} ions (SI note 8). Therefore, \ce{I} ions would jump more easily to fill a vacancy, and hence form the dominant species for diffusion via vacancies. 

A result regularly observed in simple materials is that compressive strain hinders interstitial diffusion, and tensile strain promotes it, whereas, in contrast, compressive strain promotes vacancy diffusion, and tensile strain hinders it \cite{ganster2009strain,kawamura2011self}. The general reasoning is that compression stabilizes the vacancy and destabilizes the interstitial, decreasing the diffusion barriers of the former and increasing those of the latter, whereas tension has the opposite effects. There are many exceptions to be found, but diffusion in \mixedsystem{0.5}{0.5} follows this rule. The lattice parameter of \mixedsystem{0.5}{0.5} is in between that of the pure compounds \ce{CsPbBr3} and \ce{CsPbI3}. From the perspective of \ce{Br}, \mixedsystem{0.5}{0.5} is under tensile strain, whereas from the perspective of \ce{I} it is under compressive strain. Hence, both the diffusion of Br interstitials and that of I vacancies in \mixedsystem{0.5}{0.5} should be promoted, which indeed it does considering Tables \ref{tab:diffusion-data-all} and \ref{tab:diffusion-data-CsPbIBr}.

Our results are in qualitative agreement with those obtained in experiments by McGovern \emph{et al.} on mixed MAPb(I$_x$Br$_{1-x}$)$_3$ perovskites \cite{mcgovern2021reduced}. Compared to the pure I or Br perovskites, halide migration in mixed-halide perovskites has a lower activation energy and a larger diffusion coefficient, cf. Table \ref{tab:diffusion-data-all}. In addition, in the mixed-halide perovskite one finds two sets of diffusion coefficients and activation energies, describing the migration of the Br and I ions respectively, cf. Table \ref{tab:diffusion-data-CsPbIBr}. Full quantitative agreement with experiments on MAPb(I$_x$Br$_{1-x}$)$_3$ cannot be expected, as the A cation (MA methylammonium versus Cs) is likely to have some influence on the ion migration behavior.    

\subsection{Diffusion across interfaces}

\begin{figure}
    \includegraphics{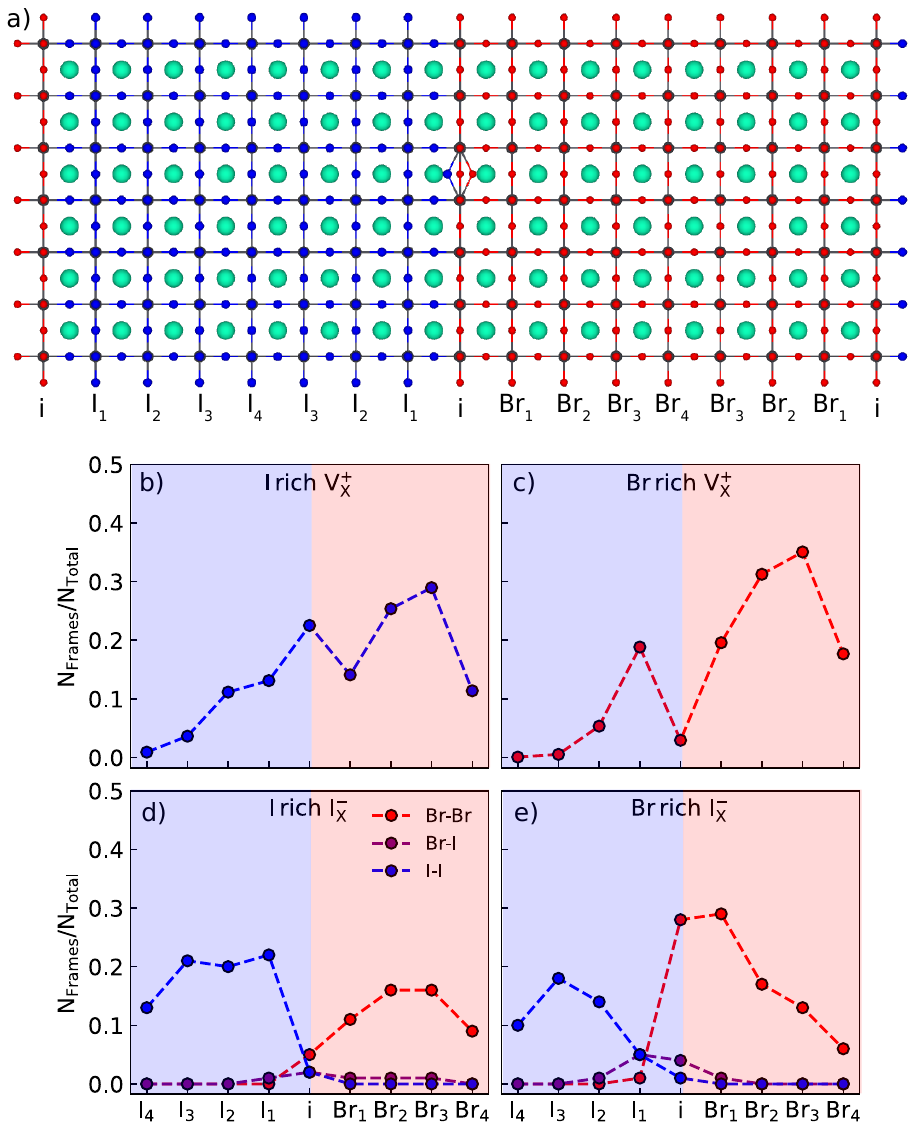}
    \caption{(a) Bulk interface and layers away from the interface in \supercell{16}{6}{6} cubic supercell of \mixedsystem{x}{1-x}, with 2 \ce{Br}-rich interfaces along the periodic boundaries. Defect occupancy at different layers ($\mathrm{N_{Frames}/N_{Total}}$) of (b) halide vacancy  for \ce{I}-rich interfaces, (c) halide vacancies for \ce{Br}-rich interfaces, (d) halide interstitials for \ce{I}-rich interfaces, and (e) halide interstitials for \ce{Br}-rich interfaces.}
    \label{fig:defect-occupancy}
\end{figure}

Besides \mixedsystem{x}{1-x} where the Br and I are {positioned randomly on halide sites} on an atomic scale, it is interesting to look at halide migration in materials comprising larger domains of pure CsPbI$_3$ and CsPbBr$_3$. This occurs, for instance, when a material in a phase-separated state is kept in a dark environment \cite{chen2021unified}. The simplest model geometry of such a state consists of layers of the pure compounds, as illustrated in Figure~\ref{fig:defect-occupancy}a, with planar interfaces along the $\langle 100 \rangle$ direction. At the interfaces between the pure I- and Br-perovskites, Pb atoms are coordinated partially by I and partially by Br atoms. 

We consider the two extreme cases, an I-rich interface where the Pb atoms are coordinated by five I atoms and one Br atom, and a Br-rich interface with Pb atoms coordinated by five Br atoms and one I atom (Figure~\ref{fig:defect-occupancy}a). The calculated difference between the static formation energies of these interfaces is very small, \SI{2.5}{meV} per Pb atom at the interface, see the SI note 9. As this is much smaller than ${k_\mathrm{B}T}$ at room temperature, we conclude that both these interfaces are equally likely to form under equilibrium conditions. 

Our model system consists of alternating layers of \ce{CsPbBr3} and \ce{CsPbI3}, each layer eight CsPbX$_3$ sublayers deep, where the interfaces are either all I-rich or all Br-rich, Figure~\ref{fig:defect-occupancy}a. We use an in-plane $6\times 6$ supercell, and start with a defect, vacancy or interstitial, positioned at a specific interface. The supercell dimensions are obtained by the procedure described in the SI note 9. As we have two different interfaces and two different defects, this gives us four different cases in total. For each case we run a 2 ns MD simulation at $T=600$ K, and repeat this five times for each system, while monitoring the position of the defect. As a result, Figure~\ref{fig:defect-occupancy}b-e shows the defect occupancies of the individual lead halide sublayers accumulated over all MD runs.

Starting with the vacancy, it clearly has a preference for residing in the \ce{CsPbBr3} layer, Figure~\ref{fig:defect-occupancy}b and c. One could map occupancies $N$ onto an energy landscape $E_{\mathrm{X}_j}$, X = Br, I and $j$ the CsPbX$_3$ sublayer index from a Boltzmann expression $N\propto \exp[-E_{\mathrm{X}_j}/k_\mathrm{B}T]$, $T=600$ K, between the two sublayers near the center of the \ce{CsPbBr3} and \ce{CsPbI3} layers one then finds an energy difference $E_{\mathrm{I}_3}-E_{\mathrm{Br}_3}>100$ meV. 

Comparing I-rich and Br-rich interfaces, there is a remarkable difference between of finding a vacancy there. Whereas the I-rich interface easily incorporates a vacancy, leading to an occupancy comparable to that of the \ce{CsPbBr3} layer, the Br-rich interface actually repels vacancies. Again, mapping onto an energy gives an interface barrier $E_\mathrm{i}\approx 130$ meV. The consequence is that I-rich interfaces are very permeable for the diffusion of vacancies, whereas Br-rich interfaces hinder the diffusion.

The interstitial gives a more complex picture. As interstitials diffuse by means of kick-out processes, they can change character, i.e., a Br interstitial can become an I interstitial in the next diffusion step and vice versa. The effects of these kick-out processes are visible in Figure~\ref{fig:defect-occupancy}d and e, where the occupancies of \ce{Pb-II-Pb} and \ce{Pb-BrBr-Pb} bridges are high in the \ce{CsPbI3} and \ce{CsPbBr3} layers, respectively, but there is no crossover of these bridges into the other layer. 

For the Br-rich interface, mixed \ce{Pb-IBr-Pb} bridges are found at or very near the interface, with negligible occupancy further from the interface. Interestingly, also in this case the I-rich interface behaves differently. It is  more permeable for I interstitials, allowing for diffusion of I into the \ce{CsPbBr3} layer, as documented by a substantial occupancy of \ce{Pb-IBr-Pb} bridges there. The general conclusion then is that Br-rich interfaces hinder inter diffusion of halide species, whereas I-rich interfaces do not. 

\begin{figure}
    \centering
    \includegraphics[]{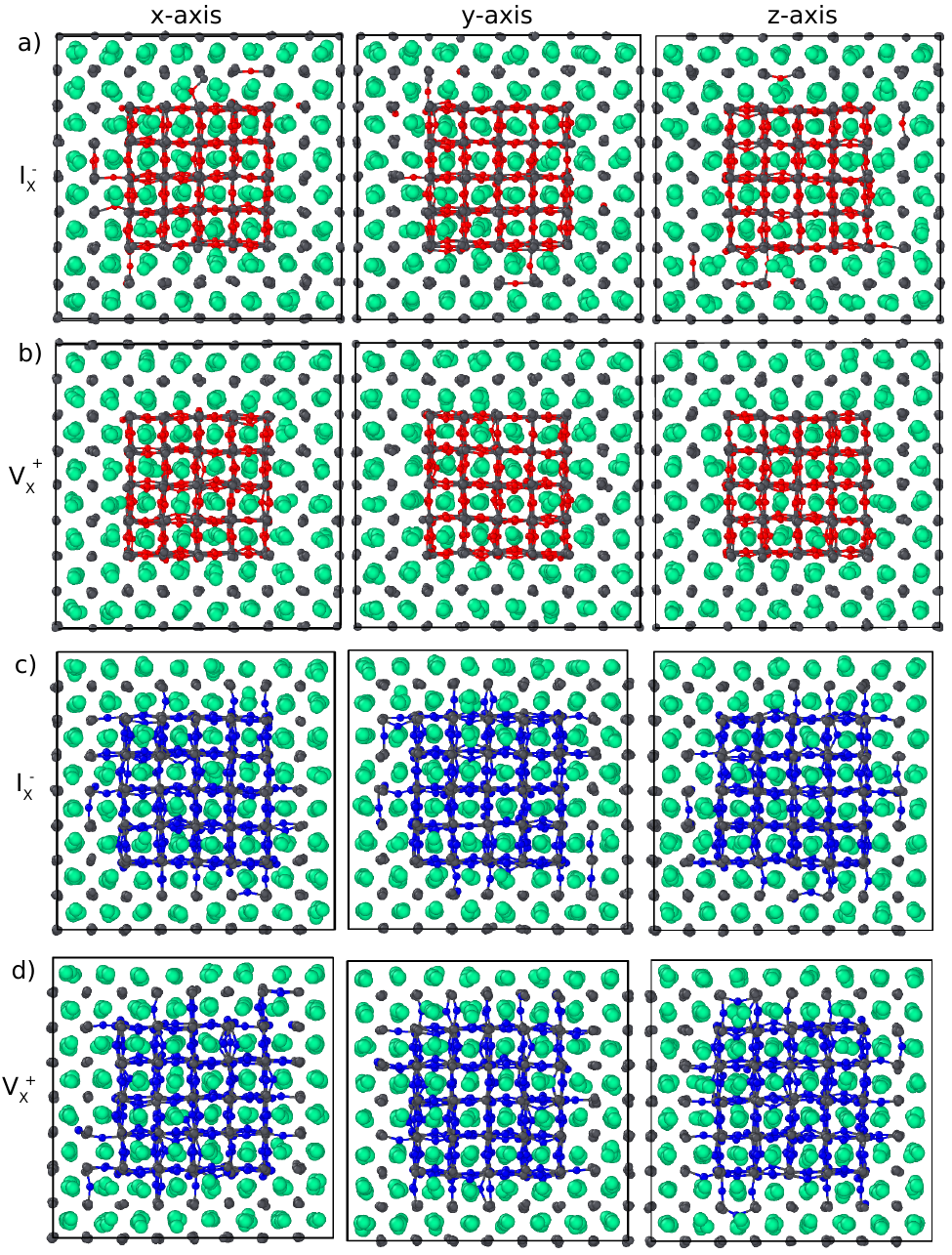}
    \caption{Final frames from the MD simulations along all three axes for (a) halide interstitial and (b) vacancy in \ce{Br} cubic domain, and (c) halide interstitial and (d) vacancy in \ce{I} cubic domain. The non-domain halide species atoms are omitted for the sake of clarity.}
    \label{fig:cubic_domain_mixing}
\end{figure}

To further demonstrate the influence of defects on halide phase mixing in \mixedsystem{x}{1-x}, we build models consisting of a $4\times 4\times 4$ cubic nanodomain of \ce{CsPbBr3} inside a \ce{CsPbI3} matrix (Figure~\ref{fig:cubic_domain_mixing}a,b), and vice versa (Figure~\ref{fig:cubic_domain_mixing}c,d). These model structures have six I-rich interfaces for I cubic nanodomains, and six Br-rich interfaces for Br cubic nanodomains. Starting with one halide point defect in the nanodomain, we perform \SI{10}{ns} long MD runs at $T=600$ K. The details of these MD runs are given in SI note 10. The final frames from the MD runs are analyzed for the number of halide atoms of the cubic nanodomain halide intermixed with one in the rest of the lattice. 

Starting with a vacancy in the \ce{CsPbBr3} nanodomain (Figure~\ref{fig:cubic_domain_mixing}b), no mixing events are observed in the 10 ns time frame. This is consistent with the interface studies discussed above, where the vacancy prefers to reside in the \ce{CsPbBr3} phase, and Br-rich interfaces hinder the diffusion. Starting with an interstitial in a \ce{CsPbBr3} nanodomain (Figure~\ref{fig:cubic_domain_mixing}a), 11 mixing events are observed, which is compatible with the fact that Br-rich interfaces are more permeable for interstitials than for vacancies.    

In contrast, the \ce{CsPbI3} nanodomain is much more fragile, with 36 mixing events for the halide vacancy in the 10 ns time frame (Figure~\ref{fig:cubic_domain_mixing}d), which is in line with the vacancy preferring the \ce{CsPbBr3} phase, and I-rich interfaces being very permeable. For the halide interstitial 29 mixing events are observed (Figure~\ref{fig:cubic_domain_mixing}b), stressing that the fragility of the \ce{CsPbI3} nanodomain does not depend too much on whether diffusion is executed by halide vacancies or interstitials.

\section{Summary and Conclusions}
In summary, we study the diffusion of halide vacancies and interstitials in \mixedsystem{x}{1-x} and CsPbI$_3$/CsPbBr$_3$ heterojunctions. We trained a neural network potential for negatively charged halide interstitial and positively charged halide vacancy defects in \mixedsystem{x}{1-x} using DFT calculations. Using this force field, we performed ns-long MD simulations to study the migration behavior of these defects and their influence on halide mixing through interfaces and domains.

We observe enhanced diffusion of both vacancies and interstitials by a factor 2-3 at room temperature in the mixed halide compounds as compared to the single halide ones, where vacancies diffuse faster than interstitials by a factor of two. Our findings furthermore suggest that the preferred migrating halide species in \mixedsystem{x}{1-x} is different for interstitial and vacancy. At room temperature, bromide migrates an order of magnitude faster than iodide for interstitials, while iodide migrates a factor of five faster than bromide for vacancies. 

These results are in qualitative agreement with experimental results on mixed MAPb(I$_x$Br$_{1-x}$)$_3$, which displays enhanced halide migration compared to MAPbI$_3$ and MAPbBr$_3$ \cite{mcgovern2021reduced}. In addition, in the mixed-halide perovskite two diffusion coefficients are found, which we suggest could describe the migration of the Br and I ions respectively. 

Diffusion across heterojunctions is governed by the interface structure, where a Br-rich interface blocks migration of vacancies in particular, and to a lesser extent for interstitials. An I-rich interface is much more permeable, both for interstitials, as well as for vacancies, and therefore facilitates diffusion and intermixing of halides. The latter is demonstrated by MD simulations on CsPbI$_3$ nanodomains in a CsPbBr$_3$ matrix and vice versa.

\begin{acknowledgement}
The authors thank Guus Bertens for his help with the compilation of LAMMPS with the \texttt{pair\_allegro} patch.  V.T. and S.T. acknowledge funding from Vidi (project no. VI.Vid.213.091) from the Dutch Research Council (NWO).
\end{acknowledgement}

% References
\bibliography{main}

@article{Kresse1996efficient,
  title = {Efficient iterative schemes for ab initio total-energy calculations using a plane-wave basis set},
  author = {Kresse, G. and Furthm\"uller, J.},
  journal = {Phys. Rev. B},
  volume = {54},
  issue = {16},
  pages = {11169--11186},
  numpages = {0},
  year = {1996},
  month = {Oct},
  publisher = {American Physical Society},
  doi = {10.1103/PhysRevB.54.11169},
  url = {https://link.aps.org/doi/10.1103/PhysRevB.54.11169}
}

@article{Jinnouchi2019phase,
  title = {Phase Transitions of Hybrid Perovskites Simulated by Machine-Learning Force Fields Trained on the Fly with Bayesian Inference},
  author = {Jinnouchi, Ryosuke and Lahnsteiner, Jonathan and Karsai, Ferenc and Kresse, Georg and Bokdam, Menno},
  journal = {Phys. Rev. Lett.},
  volume = {122},
  issue = {22},
  pages = {225701},
  numpages = {5},
  year = {2019},
  month = {Jun},
  publisher = {American Physical Society},
  doi = {10.1103/PhysRevLett.122.225701},
  url = {https://link.aps.org/doi/10.1103/PhysRevLett.122.225701}
}

@article{Jinnouchi2020descriptors,
    author = {Jinnouchi, Ryosuke and Karsai, Ferenc and Verdi, Carla and Asahi, Ryoji and Kresse, Georg},
    title = {Descriptors representing two- and three-body atomic distributions and their effects on the accuracy of machine-learned inter-atomic potentials},
    journal = {	J. Chem. Phys.},
    volume = {152},
    number = {23},
    pages = {234102},
    year = {2020},
    month = {06},
    issn = {0021-9606},
    doi = {10.1063/5.0009491},
    url = {https://doi.org/10.1063/5.0009491},
    eprint = {https://pubs.aip.org/aip/jcp/article-pdf/doi/10.1063/5.0009491/15575269/234102_1_online.pdf},
}

@article{bartok2013onrepresenting,
  title = {On representing chemical environments},
  author = {Bart\'ok, Albert P. and Kondor, Risi and Cs\'anyi, G\'abor},
  journal = {Phys. Rev. B},
  volume = {87},
  issue = {18},
  pages = {184115},
  numpages = {16},
  year = {2013},
  month = {May},
  publisher = {American Physical Society},
  doi = {10.1103/PhysRevB.87.184115},
  url = {https://link.aps.org/doi/10.1103/PhysRevB.87.184115}
}

@article{Jinnouchi2019onthefly,
  title = {On-the-fly machine learning force field generation: Application to melting points},
  author = {Jinnouchi, Ryosuke and Karsai, Ferenc and Kresse, Georg},
  journal = {Phys. Rev. B},
  volume = {100},
  issue = {1},
  pages = {014105},
  numpages = {15},
  year = {2019},
  month = {Jul},
  publisher = {American Physical Society},
  doi = {10.1103/PhysRevB.100.014105},
  url = {https://link.aps.org/doi/10.1103/PhysRevB.100.014105}
}

@article{parrinello1980crystal,
  title = {Crystal Structure and Pair Potentials: A Molecular-Dynamics Study},
  author = {Parrinello, M. and Rahman, A.},
  journal = {Phys. Rev. Lett.},
  volume = {45},
  issue = {14},
  pages = {1196--1199},
  numpages = {0},
  year = {1980},
  month = {Oct},
  publisher = {American Physical Society},
  doi = {10.1103/PhysRevLett.45.1196},
  url = {https://link.aps.org/doi/10.1103/PhysRevLett.45.1196}
}

@article{parrinello1981polymorphic,
    author = {Parrinello, M. and Rahman, A.},
    title = {Polymorphic transitions in single crystals: A new molecular dynamics method},
    journal = {J. Appl. Phys.},
    volume = {52},
    number = {12},
    pages = {7182-7190},
    year = {1981},
    month = {12},
    issn = {0021-8979},
    doi = {10.1063/1.328693},
}

@article{bartok2010gaussian,
  title = {Gaussian Approximation Potentials: The Accuracy of Quantum Mechanics, without the Electrons},
  author = {Bart\'ok, Albert P. and Payne, Mike C. and Kondor, Risi and Cs\'anyi, G\'abor},
  journal = {Phys. Rev. Lett.},
  volume = {104},
  issue = {13},
  pages = {136403},
  numpages = {4},
  year = {2010},
  month = {Apr},
  publisher = {American Physical Society},
  doi = {10.1103/PhysRevLett.104.136403},
  url = {https://link.aps.org/doi/10.1103/PhysRevLett.104.136403}
}

@article{kresse1999ultrasoft,
  title = {From ultrasoft pseudopotentials to the projector augmented-wave method},
  author = {Kresse, G. and Joubert, D.},
  journal = {Phys. Rev. B},
  volume = {59},
  issue = {3},
  pages = {1758--1775},
  numpages = {0},
  year = {1999},
  month = {Jan},
  publisher = {American Physical Society},
  doi = {10.1103/PhysRevB.59.1758},
  url = {https://link.aps.org/doi/10.1103/PhysRevB.59.1758}
}

@article{perdew1996generalized,
  title = {Generalized Gradient Approximation Made Simple},
  author = {Perdew, John P. and Burke, Kieron and Ernzerhof, Matthias},
  journal = {Phys. Rev. Lett.},
  volume = {77},
  issue = {18},
  pages = {3865--3868},
  numpages = {0},
  year = {1996},
  month = {Oct},
  publisher = {American Physical Society},
  doi = {10.1103/PhysRevLett.77.3865},
  url = {https://link.aps.org/doi/10.1103/PhysRevLett.77.3865}
}

@article{Grimme2011effect,
author = {Grimme, Stefan and Ehrlich, Stephan and Goerigk, Lars},
title = {Effect of the damping function in dispersion corrected density functional theory},
journal = {	J. Comput. Chem.},
volume = {32},
number = {7},
pages = {1456-1465},
doi = {https://doi.org/10.1002/jcc.21759},
url = {https://onlinelibrary.wiley.com/doi/abs/10.1002/jcc.21759},
eprint = {https://onlinelibrary.wiley.com/doi/pdf/10.1002/jcc.21759},
year = {2011}
}

@article{monkhorst1976special,
  title = {Special points for Brillouin-zone integrations},
  author = {Monkhorst, Hendrik J. and Pack, James D.},
  journal = {Phys. Rev. B},
  volume = {13},
  issue = {12},
  pages = {5188--5192},
  numpages = {0},
  year = {1976},
  month = {Jun},
  publisher = {American Physical Society},
  doi = {10.1103/PhysRevB.13.5188}
}

@article{tyagi_tracing_2025,
author = {Tyagi, Viren and Pols, Mike and Brocks, Geert and Tao, Shuxia},
title = {Tracing Ion Migration in Halide Perovskites with Machine Learned Force Fields},
journal = {J. Phys. Chem. Lett.},
volume = {16},
number = {20},
pages = {5153-5159},
year = {2025},
doi = {10.1021/acs.jpclett.5c01139}
}

@Article{nasstrom2020dependence,
author ="Näsström, H. and Becker, P. and Márquez, J. A. and Shargaieva, O. and Mainz, R. and Unger, E. and Unold, T.",
title  ="Dependence of phase transitions on halide ratio in inorganic \ce{CsPb(Br_{x}I_{1-x})_{3}} perovskite thin films obtained from high-throughput experimentation",
journal  ="J. Mater. Chem. A",
year  ="2020",
volume  ="8",
issue  ="43",
pages  ="22626-22631",
publisher  ="The Royal Society of Chemistry",
doi  ="10.1039/D0TA08067E"
}

@article{michaud2011mdanalysis,
author = {Michaud-Agrawal, Naveen and Denning, Elizabeth J. and Woolf, Thomas B. and Beckstein, Oliver},
title = {MDAnalysis: A toolkit for the analysis of molecular dynamics simulations},
journal = {J. Comput. Chem.},
volume = {32},
number = {10},
pages = {2319-2327},
doi = {https://doi.org/10.1002/jcc.21787},
year = {2011}
}

@article{Maginn2018best, 
title={Best Practices for Computing Transport Properties 1. Self-Diffusivity and Viscosity from Equilibrium Molecular Dynamics [Article v1.0]},
journal={LiveCoMs},
volume={1}, 
DOI={10.33011/livecoms.1.1.6324}, 
author={Maginn, Edward J. and Messerly, Richard A. and Carlson, Daniel J. and Roe, Daniel R. and Elliot, J. Richard}, 
year={2018}, 
month={Dec.}, 
pages={6324 }
}

@article{marronnier2018anharmonicity,
author = {Marronnier, Arthur and Roma, Guido and Boyer-Richard, Soline and Pedesseau, Laurent and Jancu, Jean-Marc and Bonnassieux, Yvan and Katan, Claudine and Stoumpos, Constantinos C. and Kanatzidis, Mercouri G. and Even, Jacky},
title = {Anharmonicity and Disorder in the Black Phases of Cesium Lead Iodide Used for Stable Inorganic Perovskite Solar Cells},
journal = {ACS Nano},
volume = {12},
number = {4},
pages = {3477-3486},
year = {2018},
doi = {10.1021/acsnano.8b00267},
}

@Article{thompson2022lammps,
  author = "A. P. Thompson and H. M. Aktulga and R. Berger and 
     D. S. Bolintineanu and W. M. Brown and P. S. Crozier and
     P. J. in 't Veld and A. Kohlmeyer and S. G. Moore and T. D. Nguyen and
     R. Shan and M. J. Stevens and J. Tranchida and C. Trott and S. J. Plimpton",
  title = "{LAMMPS} - a flexible simulation tool for
     particle-based materials modeling at the 
     atomic, meso, and continuum scales",
  journal = "Comp. Phys. Comm.",
  volume =  "271",
  pages =   "108171",
  year =    "2022",
  doi = "10.1016/j.cpc.2021.108171"
}

@article{Musaelian2023learning,
author={Musaelian, Albert
and Batzner, Simon
and Johansson, Anders
and Sun, Lixin
and Owen, Cameron J.
and Kornbluth, Mordechai
and Kozinsky, Boris},
title={Learning local equivariant representations for large-scale atomistic dynamics},
journal={Nat. Comm.},
year={2023},
month={Feb},
day={03},
volume={14},
number={1},
pages={579},
issn={2041-1723},
doi={10.1038/s41467-023-36329-y},
url={https://doi.org/10.1038/s41467-023-36329-y}
}

@article{Batzner2022equivariant,
author={Batzner, Simon
and Musaelian, Albert
and Sun, Lixin
and Geiger, Mario
and Mailoa, Jonathan P.
and Kornbluth, Mordechai
and Molinari, Nicola
and Smidt, Tess E.
and Kozinsky, Boris},
title={E(3)-equivariant graph neural networks for data-efficient and accurate interatomic potentials},
journal={Nat. Comm.},
year={2022},
month={May},
day={04},
volume={13},
number={1},
pages={2453},
issn={2041-1723},
doi={10.1038/s41467-022-29939-5},
url={https://doi.org/10.1038/s41467-022-29939-5}
}

@article{tang_local_2018,
	title = {Local {Observation} of {Phase} {Segregation} in {Mixed}-{Halide} {Perovskite}},
	volume = {18},
	issn = {1530-6984},
	doi = {10.1021/acs.nanolett.8b00505},
	number = {3},
	journal = {	Nano Lett.},
	author = {Tang, Xiaofeng and van den Berg, Marius and Gu, Ening and Horneber, Anke and Matt, Gebhard J. and Osvet, Andres and Meixner, Alfred J. and Zhang, Dai and Brabec, Christoph J.},
	month = mar,
	year = {2018},
	note = {Publisher: American Chemical Society},
	pages = {2172--2178},
}

@article{hoke_reversible_2015,
	title = {Reversible photo-induced trap formation in mixed-halide hybrid perovskites for photovoltaics},
	volume = {6},
	doi = {10.1039/C4SC03141E},
	number = {1},
	journal = {Chem. Sci.},
	author = {Hoke, Eric T. and Slotcavage, Daniel J. and Dohner, Emma R. and Bowring, Andrea R. and Karunadasa, Hemamala I. and McGehee, Michael D.},
	year = {2015},
	note = {Publisher: The Royal Society of Chemistry},
	pages = {613--617},
}

@article{barker_defect-assisted_2017,
	title = {Defect-{Assisted} {Photoinduced} {Halide} {Segregation} in {Mixed}-{Halide} {Perovskite} {Thin} {Films}},
	volume = {2},
	url = {https://doi.org/10.1021/acsenergylett.7b00282},
	doi = {10.1021/acsenergylett.7b00282},
	number = {6},
	urldate = {2025-02-04},
	journal = {	ACS Energy Lett.},
	author = {Barker, Alex J. and Sadhanala, Aditya and Deschler, Felix and Gandini, Marina and Senanayak, Satyaprasad P. and Pearce, Phoebe M. and Mosconi, Edoardo and Pearson, Andrew J. and Wu, Yue and Srimath Kandada, Ajay Ram and Leijtens, Tomas and De Angelis, Filippo and Dutton, Siân E. and Petrozza, Annamaria and Friend, Richard H.},
	month = jun,
	year = {2017},
	note = {Publisher: American Chemical Society},
	pages = {1416--1424},
}

@article{chen2021unified,
  title={Unified theory for light-induced halide segregation in mixed halide perovskites},
  author={Chen, Zehua and Brocks, Geert and Tao, Shuxia and Bobbert, Peter A},
  journal={Nat. Commun.},
  volume={12},
  number={1},
  pages={2687},
  year={2021}
}

@article{chen2022light,
  title = {Light-tunable three-phase coexistence in mixed halide perovskites},
  author = {Chen, Zehua and Brocks, Geert and Tao, Shuxia and Bobbert, Peter A.},
  journal = {Phys. Rev. B},
  volume = {106},
  issue = {13},
  pages = {134110},
  numpages = {11},
  year = {2022},
  month = {Oct},
  doi = {10.1103/PhysRevB.106.134110}
}

@article{Xue2022intrinsic,
  title = {Intrinsic defects in primary halide perovskites: A first-principles study of the thermodynamic trends},
  author = {Xue, Haibo and Brocks, Geert and Tao, Shuxia},
  journal = {Phys. Rev. Mater.},
  volume = {6},
  issue = {5},
  pages = {055402},
  numpages = {11},
  year = {2022},
  month = {May},
  publisher = {American Physical Society},
  doi = {10.1103/PhysRevMaterials.6.055402}
}

@article{Xue2021first,
  title = {First-principles calculations of defects in metal halide perovskites: A performance comparison of density functionals},
  author = {Xue, Haibo and Brocks, Geert and Tao, Shuxia},
  journal = {Phys. Rev. Mater.},
  volume = {5},
  issue = {12},
  pages = {125408},
  numpages = {11},
  year = {2021},
  month = {Dec},
  publisher = {American Physical Society},
  doi = {10.1103/PhysRevMaterials.5.125408}
}

@article{Wang2018defects,
author={Wang, Feng
and Bai, Sai
and Tress, Wolfgang
and Hagfeldt, Anders
and Gao, Feng},
title={Defects engineering for high-performance perovskite solar cells},
journal={npj flex. electron.},
year={2018},
month={Aug},
day={14},
volume={2},
number={1},
pages={22},
issn={2397-4621},
doi={10.1038/s41528-018-0035-z}
}

@article{Ball2016defects,
author={Ball, James M.
and Petrozza, Annamaria},
title={Defects in perovskite-halides and their effects in solar cells},
journal={Nat. Energy},
year={2016},
month={Oct},
day={31},
volume={1},
number={11},
pages={16149},
issn={2058-7546},
doi={10.1038/nenergy.2016.149}
}

@article{eames_ionic_2015,
	title = {Ionic transport in hybrid lead iodide perovskite solar cells},
	volume = {6},
	copyright = {2015 The Author(s)},
	issn = {2041-1723},
	doi = {10.1038/ncomms8497},
	language = {en},
	number = {1},
	urldate = {2024-07-18},
	journal = {Nat. Commun.},
	author = {Eames, Christopher and Frost, Jarvist M. and Barnes, Piers R. F. and O’Regan, Brian C. and Walsh, Aron and Islam, M. Saiful},
	month = jun,
	year = {2015},
	note = {Publisher: Nature Publishing Group},
	pages = {7497},
}

@article{sarkar2024effects,
author = {Sarkar, Gourab and Ghosh, Dibyajyoti},
title = {Effects of Lattice Compression on Halogen Ion Diffusion Dynamics in Mixed Halide Perovskites},
journal = {ACS Appl. Energy Mater.},
volume = {7},
number = {15},
pages = {6376-6383},
year = {2024},
doi = {10.1021/acsaem.4c01063}

}

@article{liang2023role,
author = {Liang, Xiao and Duan, Dawei and Al-Handawi, Marieh B. and Wang, Fei and Zhou, Xianfang and Ge, Chuang-ye and Lin, Haoran and Zhu, Quanyao and Li, Liang and Naumov, Panče and Hu, Hanlin},
title = {The Role of Ionic Liquids in Performance Enhancement of Two-Step Perovskite Photovoltaics},
journal = {Sol. RRL},
volume = {7},
number = {1},
pages = {2200856},
doi = {https://doi.org/10.1002/solr.202200856},
year = {2023}
}

@article{Lin2018perovskite,
author={Lin, Kebin
and Xing, Jun
and Quan, Li Na
and de Arquer, F. Pelayo Garc{\'i}a
and Gong, Xiwen
and Lu, Jianxun
and Xie, Liqiang
and Zhao, Weijie
and Zhang, Di
and Yan, Chuanzhong
and Li, Wenqiang
and Liu, Xinyi
and Lu, Yan
and Kirman, Jeffrey
and Sargent, Edward H.
and Xiong, Qihua
and Wei, Zhanhua},
title={Perovskite light-emitting diodes with external quantum efficiency exceeding 20 per cent},
journal={Nature},
year={2018},
month={Oct},
day={01},
volume={562},
number={7726},
pages={245-248},
issn={1476-4687},
doi={10.1038/s41586-018-0575-3}
}

@article{Liu2019low,
title = {Low-temperature-gradient crystallization for multi-inch high-quality perovskite single crystals for record performance photodetectors},
journal = {	Mater. Today},
volume = {22},
pages = {67-75},
year = {2019},
issn = {1369-7021},
doi = {https://doi.org/10.1016/j.mattod.2018.04.002},
author = {Yucheng Liu and Yunxia Zhang and Zhou Yang and Jiangshan Feng and Zhuo Xu and Qingxian Li and Mingxin Hu and Haochen Ye and Xu Zhang and Ming Liu and Kui Zhao and Shengzhong(Frank) Liu}
}

@inproceedings{schutt2017schnet,
 author = {Sch\"{u}tt, Kristof and Kindermans, Pieter-Jan and Sauceda Felix, Huziel Enoc and Chmiela, Stefan and Tkatchenko, Alexandre and M\"{u}ller, Klaus-Robert},
 booktitle = {Advances in Neural Information Processing Systems},
 editor = {I. Guyon and U. Von Luxburg and S. Bengio and H. Wallach and R. Fergus and S. Vishwanathan and R. Garnett},
 pages = {},
 publisher = {Curran Associates, Inc.},
 title = {SchNet: A continuous-filter convolutional neural network for modeling quantum interactions},
 url = {https://proceedings.neurips.cc/paper_files/paper/2017/file/303ed4c69846ab36c2904d3ba8573050-Paper.pdf},
 volume = {30},
 year = {2017}
}

@article{Geiger2022e3nn,
  doi = {10.48550/ARXIV.2207.09453},
  author = {Mario Geiger and Tess Smidt},
  title = {e3nn: Euclidean Neural Networks},
  publisher = {Cornell University},
  year = {2022},
  eprint       = {arXiv:2207.09453},
  pubstate      = {\bibstring{prepublished}},
  note = {Preprint at https://doi.org/10.48550/arXiv.2207.09453}
}

@article{Henkelman2020climbing,
    author = {Henkelman, Graeme and Uberuaga, Blas P. and Jónsson, Hannes},
    title = {A climbing image nudged elastic band method for finding saddle points and minimum energy paths},
    journal = {	J. Chem. Phys.},
    volume = {113},
    number = {22},
    pages = {9901-9904},
    year = {2000},
    month = {12},
    issn = {0021-9606},
    doi = {10.1063/1.1329672}
}

@article{ganster2009strain,
  title = {Strain effect on self-diffusion in silicon: Numerical study},
  author = {Ganster, P. and Tr\'eglia, G. and Sa\'ul, A.},
  journal = {Phys. Rev. B},
  volume = {79},
  issue = {11},
  pages = {115205},
  numpages = {11},
  year = {2009},
  month = {Mar},
  publisher = {American Physical Society},
  doi = {10.1103/PhysRevB.79.115205}
}

@article{kawamura2011self,
    author = {Kawamura, Yoko and Uematsu, Masashi and Hoshi, Yusuke and Sawano, Kentarou and Myronov, Maksym and Shiraki, Yasuhiro and Haller, Eugene E. and Itoh, Kohei M.},
    title = {Self-diffusion in compressively strained Ge},
    journal = {	J. Appl. Phys.},
    volume = {110},
    number = {3},
    pages = {034906},
    year = {2011},
    month = {08},
    issn = {0021-8979},
    doi = {10.1063/1.3608171}
}

@article{mcgovern2021reduced,
author = {McGovern, Lucie and Grimaldi, Gianluca and Futscher, Moritz H. and Hutter, Eline M. and Muscarella, Loreta A. and Schmidt, Moritz C. and Ehrler, Bruno},
title = {Reduced Barrier for Ion Migration in Mixed-Halide Perovskites},
journal = {	ACS Appl. Energy Mater.},
volume = {4},
number = {12},
pages = {13431-13437},
year = {2021},
doi = {10.1021/acsaem.1c03095}

}

@article{Yi2016entropic,
author ="Yi, Chenyi and Luo, Jingshan and Meloni, Simone and Boziki, Ariadni and Ashari-Astani, Negar and Grätzel, Carole and Zakeeruddin, Shaik M. and Röthlisberger, Ursula and Grätzel, Michael",
title  ="Entropic stabilization of mixed A-cation \ce{ABX_3} metal halide perovskites for high performance perovskite solar cells",
journal  ="Energy Environ. Sci.",
year  ="2016",
volume  ="9",
issue  ="2",
pages  ="656-662",
publisher  ="The Royal Society of Chemistry",
doi  ="10.1039/C5EE03255E"
}

@article{saliba2016cesium,
author ="Saliba, Michael and Matsui, Taisuke and Seo, Ji-Youn and Domanski, Konrad and Correa-Baena, Juan-Pablo and Nazeeruddin, Mohammad Khaja and Zakeeruddin, Shaik M. and Tress, Wolfgang and Abate, Antonio and Hagfeldt, Anders and Grätzel, Michael",
title  ="Cesium-containing triple cation perovskite solar cells: improved stability{,} reproducibility and high efficiency",
journal  ="Energy Environ. Sci.",
year  ="2016",
volume  ="9",
issue  ="6",
pages  ="1989-1997",
publisher  ="The Royal Society of Chemistry",
doi  ="10.1039/C5EE03874J"}

@article{Jung2017influence,
author = {Jung, Young-Kwang and Lee, Ji-Hwan and Walsh, Aron and Soon, Aloysius},
title = {Influence of Rb/Cs Cation-Exchange on Inorganic Sn Halide Perovskites: From Chemical Structure to Physical Properties},
journal = {	Chem. Mater.},
volume = {29},
number = {7},
pages = {3181-3188},
year = {2017},
doi = {10.1021/acs.chemmater.7b00260},
    note ={PMID: 28435185},

}

@article{Draguta2017rationalizing,
author={Draguta, Sergiu
and Sharia, Onise
and Yoon, Seog Joon
and Brennan, Michael C.
and Morozov, Yurii V.
and Manser, Joseph S.
and Kamat, Prashant V.
and Schneider, William F.
and Kuno, Masaru},
title={Rationalizing the light-induced phase separation of mixed halide organic--inorganic perovskites},
journal={Nat. Commun.},
year={2017},
month={Aug},
day={04},
volume={8},
number={1},
pages={200},
issn={2041-1723},
doi={10.1038/s41467-017-00284-2}
}

@article{Wang2019suppressed,
author={Wang, Xi
and Ling, Yichuan
and Lian, Xiujun
and Xin, Yan
and Dhungana, Kamal B.
and Perez-Orive, Fernando
and Knox, Javon
and Chen, Zhizhong
and Zhou, Yan
and Beery, Drake
and Hanson, Kenneth
and Shi, Jian
and Lin, Shangchao
and Gao, Hanwei},
title={Suppressed phase separation of mixed-halide perovskites confined in endotaxial matrices},
journal={Nat. Commun.},
year={2019},
month={Feb},
day={11},
volume={10},
number={1},
pages={695},
issn={2041-1723},
doi={10.1038/s41467-019-08610-6}
}

@article{yunative2016,
	title = {Native {Defect}-{Induced} {Hysteresis} {Behavior} in {Organolead} {Iodide} {Perovskite} {Solar} {Cells}},
	volume = {26},
	issn = {1616-3028},
	url = {https://onlinelibrary.wiley.com/doi/abs/10.1002/adfm.201504997},
	doi = {10.1002/adfm.201504997},
	language = {en},
	number = {9},
	urldate = {2025-12-09},
	journal = {	Adv. Funct. Mater.},
	author = {Yu, Hui and Lu, Haipeng and Xie, Fangyan and Zhou, Shuang and Zhao, Ni},
	year = {2016},
	pages = {1411--1419},
}

@article{liiodine2016,
	title = {Iodine {Migration} and its {Effect} on {Hysteresis} in {Perovskite} {Solar} {Cells}},
	volume = {28},
	issn = {1521-4095},
	doi = {10.1002/adma.201503832},
	number = {12},
	urldate = {2025-12-09},
	journal = {Adv. Mater.},
	author = {Li, Cheng and Tscheuschner, Steffen and Paulus, Fabian and Hopkinson, Paul E. and Kießling, Johannes and Köhler, Anna and Vaynzof, Yana and Huettner, Sven},
	year = {2016},
	pages = {2446--2454},
}

@article{game2017ions,
	title = {Ions {Matter}: {Description} of the {Anomalous} {Electronic} {Behavior} in {Methylammonium} {Lead} {Halide} {Perovskite} {Devices}},
	volume = {27},
	copyright = {© 2017 WILEY-VCH Verlag GmbH \& Co. KGaA, Weinheim},
	issn = {1616-3028},
	shorttitle = {Ions {Matter}},
	doi = {10.1002/adfm.201606584},
	language = {en},
	number = {16},
	urldate = {2025-12-09},
	journal = {	Adv. Funct. Mater.},
	author = {Game, Onkar S. and Buchsbaum, Gabriel J. and Zhou, Yuanyuan and Padture, Nitin P. and Kingon, Angus I.},
	year = {2017},
	pages = {1606584},
}

@article{schmidt2026quantification,
author = {Schmidt, Moritz C. and Alvarez, Agustin O. and Pallotta, Riccardo and Seid, Biruk A. and de Boer, Jeroen J. and Thiesbrummel, Jarla and Lang, Felix and Grancini, Giulia and Ehrler, Bruno},
title = {Quantification of Mobile Ions in Perovskite Solar Cells with Thermally Activated Ion Current Measurements},
journal = {	ACS Energy Lett.},
volume = {11},
number = {1},
pages = {409-418},
year = {2026},
doi = {10.1021/acsenergylett.5c02224},

}

@article{mosconi2016light,
	title = {Light-induced annihilation of {Frenkel} defects in organo-lead halide perovskites},
	volume = {9},
	issn = {1754-5692, 1754-5706},
	doi = {10.1039/C6EE01504B},
	language = {en},
	number = {10},
	urldate = {2025-12-09},
	journal = {Energy Environ. Sci.},
	author = {Mosconi, Edoardo and Meggiolaro, Daniele and Snaith, Henry J. and Stranks, Samuel D. and De Angelis, Filippo},
	year = {2016},
	pages = {3180--3187},
}

@article{behler2021four,
author = {Behler, J{\"o}rg},
title = {Four Generations of High-Dimensional Neural Network Potentials},
journal = {	Chem. Rev.},
volume = {121},
number = {16},
pages = {10037-10072},
year = {2021},
doi = {10.1021/acs.chemrev.0c00868},
    note ={PMID: 33779150},

}

@article{Wu2023applications,
author = {Wu, Shiru and Yang, Xiaowei and Zhao, Xun and Li, Zhipu and Lu, Min and Xie, Xiaoji and Yan, Jiaxu},
title = {Applications and Advances in Machine Learning Force Fields},
journal = {	J. Chem. Inf. Model.},
volume = {63},
number = {22},
pages = {6972-6985},
year = {2023},
doi = {10.1021/acs.jcim.3c00889},
    note ={PMID: 37751546},

}

@article{Unke2021machine,
author = {Unke, Oliver T. and Chmiela, Stefan and Sauceda, Huziel E. and Gastegger, Michael and Poltavsky, Igor and Sch{\"u}tt, Kristof T. and Tkatchenko, Alexandre and M{\"u}ller, Klaus-Robert},
title = {Machine Learning Force Fields},
journal = {	Chem. Rev.},
volume = {121},
number = {16},
pages = {10142-10186},
year = {2021},
doi = {10.1021/acs.chemrev.0c01111},
    note ={PMID: 33705118},

}

@article{Sarkar2024ion,
author = {Sarkar, Gourab and Deswal, Priyanka and Ghosh, Dibyajyoti},
title = {Ion Diffusion Dynamics and Halogen Mixing at the Heterojunction of Halide Perovskites: Atomistic Insights},
journal = {J. Phys. Chem. C},
volume = {128},
number = {4},
pages = {1762-1772},
year = {2024},
doi = {10.1021/acs.jpcc.3c06329},

}

@article{Meggiolaro2018first,
author = {Meggiolaro, Daniele and De Angelis, Filippo},
title = {First-Principles Modeling of Defects in Lead Halide Perovskites: Best Practices and Open Issues},
journal = {	ACS Energy Lett.},
volume = {3},
number = {9},
pages = {2206-2222},
year = {2018},
doi = {10.1021/acsenergylett.8b01212},

}

@article{mosquera-lois2025point,
author ="Mosquera-Lois, Irea and Klarbring, Johan and Walsh, Aron",
title  ="Point defect formation at finite temperatures with machine learning force fields",
journal  ="Chem. Sci.",
year  ="2025",
volume  ="16",
issue  ="20",
pages  ="8878-8888",
publisher  ="The Royal Society of Chemistry",
doi  ="10.1039/D4SC08582E",
}

@article{kessels2025tailoring,
author = {Kessels, Lana M. and Remmerswaal, Willemijn H. M. and Schipper, Nick R. M. and Bellini, Laura and Kwan, Henry and Wienk, Martijn M. and Janssen, René A. J.},
title = {Tailoring the Crystallization Behavior of Mixed Lead-Tin Mixed-Halide Perovskites for Optimal-Bandgap Solar Cells},
journal = {	Adv. Sci.},
year  ={2025},
volume = {n/a},
number = {n/a},
pages = {e20948},
doi = {https://doi.org/10.1002/advs.202520948},
}

@article{liang2022selective,
author = {Liang, Zheng and Xu, Huifen and Zhang, Yong and Liu, Guozhen and Chu, Shenglong and Tao, Yuli and Xu, Xiaoxiao and Xu, Shendong and Zhang, Liying and Chen, Xiaojing and Xu, Baomin and Xiao, Zhengguo and Pan, Xu and Ye, Jiajiu},
title = {A Selective Targeting Anchor Strategy Affords Efficient and Stable Ideal-Bandgap Perovskite Solar Cells},
journal = {Adv. Mater.},
volume = {34},
number = {18},
pages = {2110241},
doi = {https://doi.org/10.1002/adma.202110241},
year = {2022}
}

@article{lahnsteiner2018finite,
  title = {Finite-temperature structure of the ${\mathrm{MAPbI}}_{3}$ perovskite: Comparing density functional approximations and force fields to experiment},
  author = {Lahnsteiner, Jonathan and Kresse, Georg and Heinen, Jurn and Bokdam, Menno},
  journal = {Phys. Rev. Mater.},
  volume = {2},
  issue = {7},
  pages = {073604},
  numpages = {14},
  year = {2018},
  month = {Jul},
  publisher = {American Physical Society},
  doi = {10.1103/PhysRevMaterials.2.073604}
}

\end{document}

\clearpage

\tableofcontents

\clearpage
\section{Defect migration in \ce{CsPbBr_{3}}}\label{CsPbBr3}
\subsection{Model training}\label{CsPbBr3_training}
We start by training different MLFFs for all three negative, neutral, and positive charge states of halide interstitial and halide vacancy defects in \ce{CsPbBr3}. Variants of the GAP-SOAP model\cite{bartok2013onrepresenting,bartok2010gaussian} are trained as implemented in VASP\cite{Kresse1996efficient,Jinnouchi2019onthefly,Jinnouchi2019phase}, where the training structures are sampled from three MD runs. These runs are performed at different temperatures for each system to ensure that the force fields can describe diverse atomic environments during the production runs. Temperatures for these runs are given in Table~\ref{tab:table1}. These runs are performed using \supercell{2}{2}{2} cubic supercells (8 units of \ce{CsPbBr_{3}}) with one halide point defect for \SI{100}{ps} with MD time-steps of \SI{2}{fs}. Most of these runs are performed in \textit{NpT} ensembles at $\mathrm{10^{5}}$ \si{Pa} pressure. To maintain constant temperature and pressure, Parinello-Rahman dynamics is used \cite{parrinello1980crystal,parrinello1981polymorphic}. For \chargeddefect{V}{-}, unphysically large volume expansions are observed during the first training run (\SI{700}{K}), possibly due to excess electronic charge in a relatively small training structure. Hence, the volume is kept constant during this run, while the shape of the cell is allowed to change.

During these training runs, the frames are evaluated using density functional theory (DFT) calculations whenever the estimation error in forces exceeds a threshold value. For these DFT calculations, the projector-augmented wave (PAW) \cite{kresse1999ultrasoft} technique is used to model the electron-ion interaction, with the outermost electrons of \ce{Br} ($\mathrm{4s^{2}4p^{5}}$), \ce{Cs} ($\mathrm{5s^{2}5p^{6}6s^{1}}$), and \ce{Pb} ($\mathrm{6s^{2}6p^{2}}$) treated as valence electrons. The electronic interactions are modeled using the PBE exchange-correlation functional within the generalized gradient approximation (GGA) supplemented with D3-BJ Vanderwaals contributions \cite{perdew1996generalized,Grimme2011effect}. Energy convergence criterion of $10^{-6}$\SI{}{eV} is used for the Self-consistent field cycles. The calculations are performed using \supercell{2}{2}{2} Monkhorst-Pack k-grid \cite{monkhorst1976special} and a kinetic energy cutoff of \SI{300}{eV}.

\begin{table}[]
    \centering
    \begin{tabular}{|c c c c|}
        \hline
        System & Step 1 & Step 2 & Step 3 \\
        \hline
        \chargeddefect{I}{-} & \SI{700}{K} & \SI{750}{K} & \SI{600}{K} \\
        \chargeddefect{I}{0} & \SI{700}{K} & \SI{750}{K} & \SI{600}{K} \\
        \chargeddefect{I}{+} & \SI{700}{K} & \SI{750}{K} & \SI{600}{K} \\
        \chargeddefect{V}{-} & \SI{700}{K}\textsuperscript{*} & \SI{750}{K} & \SI{750}{K} \\
        \chargeddefect{V}{0} & \SI{750}{K} & \SI{750}{K} & \SI{800}{K} \\
        \chargeddefect{V}{+} & \SI{700}{K} & \SI{750}{K} & \SI{700}{K} \\
        \hline
    \end{tabular}
    \caption{\textmd{Temperatures at which training was performed for each defect system. Here \textsuperscript{*} corresponds to the runs for which the cell volume was kept constant.}}
    \label{tab:table1}
\end{table}

To describe the local chemical environments, a combination of two descriptors similar to the smooth overlap of atomic positions (SOAP) descriptor is used \cite{bartok2013onrepresenting}. The first of these is a descriptor closely resembling a radial distribution function that can also be regarded as a two-body descriptor, for which a cutoff $\rho_{i}^{(2)}$ of $7\,\text{\AA}$ is used. The second descriptor is purely angular, containing no two-body components, for which a cutoff $\rho_{i}^{(3)}$ of $6\,\text{\AA}$ is used \cite{Jinnouchi2020descriptors}. The atomic positions are expanded using Gaussian distributions of width $0.5\,\text{\AA}$. The radial descriptor is expanded using 12 radial basis functions, whereas the angular descriptor is expanded using 8 radial basis functions and spherical harmonics with a maximum angular momentum quantum number $L_{\max}$ of 6. As the structures are sampled, a set of local reference configurations is constructed. The maximum size of this set is 4500 for all systems. The local potential energy of an atom in a structure from the training set is expressed as a linear combination of Gaussian kernels that measure the similarity between the local reference configuration from the training set and the basis set. A polynomial power of 4 is used for these kernels, and the weight of the radial descriptor in these kernels is set to 0.1, setting the weight of the angular descriptor to 0.9.

\begin{figure}
    \centering
    \includegraphics{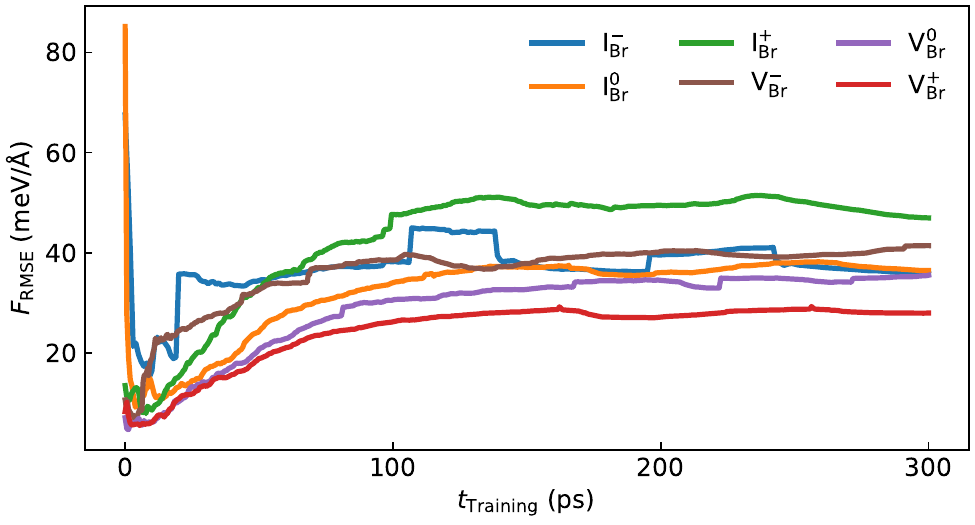}
    \caption{Root mean squared error in forces ($F_{\mathrm{RMSE}}$) calculated by the force field compared to DFT calculated forces during training of all charge states of bromide interstitial and vacancy defects in \ce{CsPbBr3}.}
    \label{fig:rmse_forces_CsPbBr3}
\end{figure}

\begin{figure}
    \centering
    \includegraphics{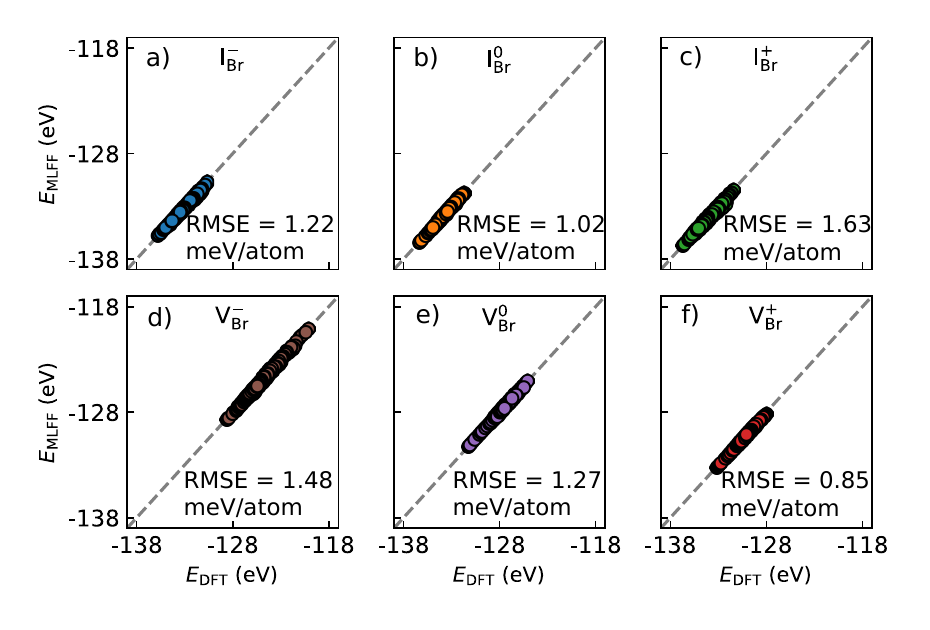}
    \caption{Comparison between energies calculated using DFT with energies calculated using the final MLFF for all structures in the training set of all defect systems of \ce{CsPbBr3}, along with the corresponding root mean squared errors.}
    \label{fig:final-training-energies-CsPbBr3}
\end{figure}

The forces and energies predicted by the current force field can be compared with forces and energies calculated using DFT during training. The root mean squared difference between these forces over training time is given in Figure ~\ref{fig:rmse_forces_CsPbBr3}. These errors plateau over time for all systems, signifying that the force field accurately captures the potential energy surface of defective perovskite systems, making it suitable for studying their dynamics. Furthermore, the energies were calculated for all training structures using the final force fields and compared with DFT-calculated energies. These comparisons are given in Figure~\ref {fig:final-training-energies-CsPbBr3}. The root mean squared errors from these comparisons ($E_{\mathrm{RMSE}}$) are in the order \SI{1}{meV/atom}, as the energies of the training structures are of the order of \SI{1}{eV/atom}. These values indicate that the force fields are accurately learning the energies during training.

Finally, all the force fields are refitted onto faster descriptors without any error estimate to speed up their evaluations for use in production runs. The number of structures in the training sets ($N_{\mathrm{DFT}}$) and the number of local reference configurations for the different elements in the MLFFs ($N_{\mathrm{basis}}$) are given in Table~\ref{tab:table2}

\begin{table}[]
    \centering
    \begin{tabular}{|c c c c c|}
        \hline
        {System} & {$N_{\mathrm{DFT}}(-)$} & \multicolumn{3}{c|}{$N_{\mathrm{basis}}(-)$} \\
        \multicolumn{2}{|c}{} & {Cs} & {Pb} & {Br} \\
        \hline
        \chargeddefect{I}{-} & {1328} & {1723} & {1764} & {4129} \\
        \chargeddefect{I}{0} & {2057} & {1449} & {1903} & {4002} \\
        \chargeddefect{I}{+} & {2370} & {1561} & {1841} & {4500} \\
        \chargeddefect{V}{-} & {3317} & {787} & {2986} & {4500} \\
        \chargeddefect{V}{0} & {1965} & {2039} & {2167} & {4500} \\
        \chargeddefect{V}{+} & {2124} & {1615} & {2202} & {4500} \\
        \hline
    \end{tabular}
    \caption{\textmd{{Size of the training set and number of local reference configurations in the MLFFs for each defect system.}}}
    \label{tab:table2}
\end{table}

\clearpage

\subsection{Model validation}\label{CsPbBr3_validation}
\begin{figure}
    \centering
    \includegraphics{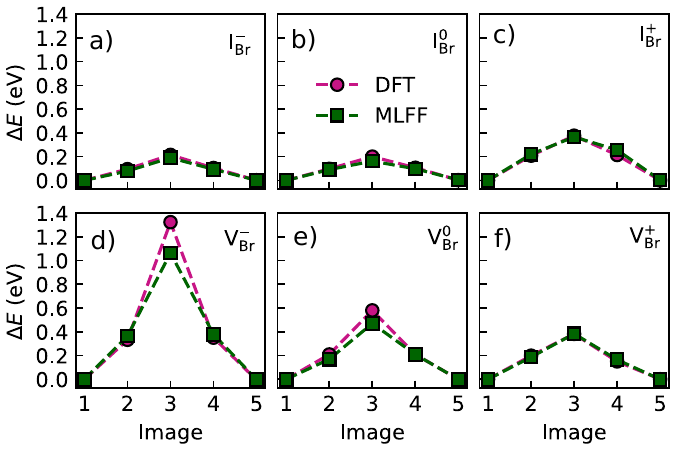}
    \caption{Energies along the NEB migration paths for different charge states of the (a-c) \ce{Br} interstitial and (d-f) \ce{Br} vacancy defects in \ce{CsPbBr3}, calculated using DFT and the MLFFs. The energy minima are set to 0.}
    \label{fig:CsPbBr3_NEB_energies}
\end{figure}

To test the accuracy of these models, we compare the energies predicted by the force fields with those calculated using DFT for migration paths as they would enter transition state theory (TST). These paths are constructed using climbing image nudged elastic band (CI-NEB) calculations on \supercell{2}{2}{1} orthorhombic supercells (16 units of \ce{CsPbBr_{3}}) with one point defect. Three intermediate images, connected by springs with spring constant $\mathrm{5\,eV/\text{\AA}^{2}}$ are optimized using kinetic energy cut-off of \SI{300}{eV} and \supercell{2}{2}{3} Monkhorst-Pack \textit{k}-point grid. The computed energies are given in Figure~\ref {fig:CsPbBr3_NEB_energies}, and the migration barriers are given in Table~\ref{tab:table3}. The DFT calculated trends in migration barriers are well captured by the MLFFs (Table ~\ref{tab:table3}). Further, the difference between the migration barriers calculated using the MLFFs and DFT are within \SI{0.11}{eV} for all defects except \chargeddefect{V}{-}, where the difference is \SI{0.26}{eV} (Figure ~\ref{fig:CsPbBr3_NEB_energies}d). This could be due to only a few migration events being sampled while training this MLFF.

\begin{table}[]
    \centering
    \begin{tabular}{|c c c|}
        \hline
        System & $E_{\mathrm{b}}^{\mathrm{DFT}}$ (eV) & $E_{\mathrm{b}}^{\mathrm{MLFF}}$ (eV) \\
        \hline
        \chargeddefect{I}{-} & {0.21} & {0.19} \\
        \chargeddefect{I}{0} & {0.20} & {0.16} \\
        \chargeddefect{I}{+} & {0.38} & {0.37} \\
        \chargeddefect{V}{-} & {1.32} & {1.06} \\
        \chargeddefect{V}{0} & {0.58} & {0.47} \\
        \chargeddefect{V}{+} & {0.38} & {0.38} \\
        \hline
    \end{tabular}
    \caption{\textmd{Defect migration barriers ($\mathrm{E_{b}}$) calculated using DFT/TST/CI-NEB migration paths.}}
    \label{tab:table3}
\end{table}

\begin{figure}
    \centering
    \includegraphics{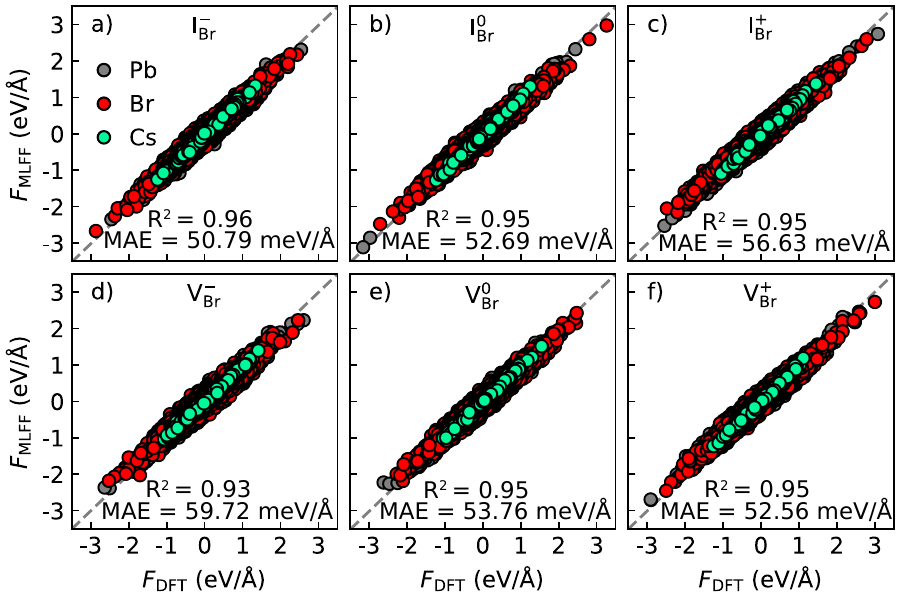}
    \caption{Comparison between forces acting on all atoms calculated by the MLFFs with forces calculated using DFT for different charge states of halide interstitial (a-c) and halide vacancy (d-f) defects in \supercell{6}{6}{6} supercells of \ce{CsPbBr3}, along with corresponding $\mathrm{R^{2}}$ values and mean absolute errors (MAE).}
    \label{fig:CsPbBr_forces_all}
\end{figure}

\begin{figure}
    \centering
    \includegraphics{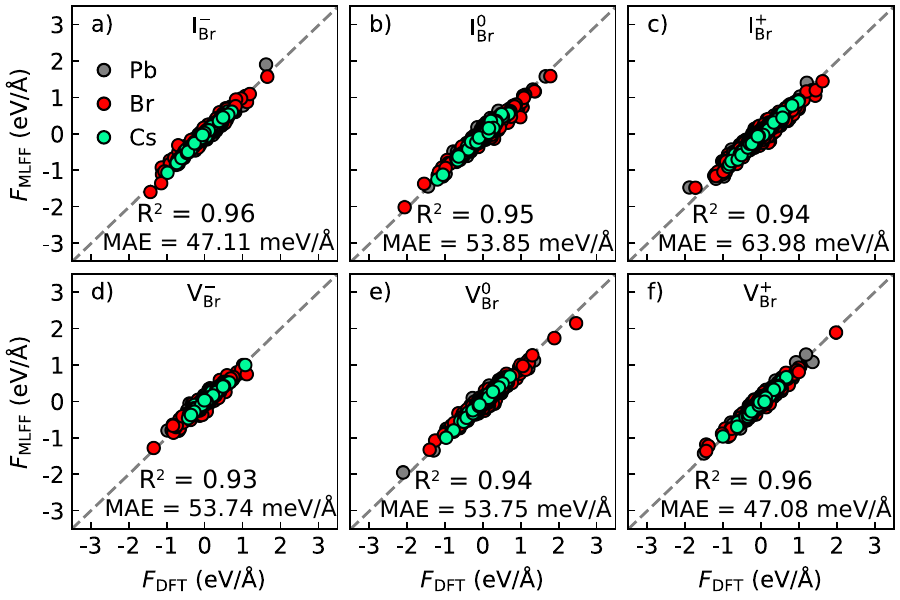}
    \caption{Comparison between forces acting on atoms in the defect environment calculated by the MLFFs with forces calculated using DFT for different charge states of halide interstitial (a-c) and halide vacancy (d-f) defects in \supercell{6}{6}{6} supercells of \ce{CsPbBr3}, along with corresponding $\mathrm{R^{2}}$ values and mean absolute errors (MAE).}
    \label{fig:CsPbBr_forces_defect}
\end{figure}

To test the accuracy of the MLFFs during the MD production runs, we sample {20} structures from \SI{0.5}{\ns} MD runs at \SI{600}{\K} performed using \supercell{6}{6}{6} cubic supercells (216 units of \ce{CsPbBr_{3}}) with one bromide point defect, and compare forces calculated using DFT with forces calculated using the force fields for these structures. We compare forces acting on all atoms (Figure~\ref{fig:CsPbBr_forces_all}) and also, in particular, the forces of atoms close to the defect (Figure~\ref{fig:CsPbBr_forces_defect}) to check the accuracy with which the force fields could describe the defect environments. The procedure for identifying atoms in the defect environment is explained in the SI note 7 of our previous work \cite{tyagi_tracing_2025}. From these comparisons, we note that the $\mathrm{R^{2}}\,\geq\,0.93$ and $\mathrm{MAE\,\leq\,\SI{59.72}{meV/\angstrom}}$ for forces acting on all atoms, and $\mathrm{R^{2}}\,\geq\,0.93$ and $\mathrm{MAE\,\leq\,\SI{63.98}{meV/\angstrom}}$ for forces acting on atoms in the defect environment. As the forces acting on the atoms are of the order of $\SI{1}{eV/\angstrom}$, all force fields are transferable to bigger system sizes and accurate in capturing the defect environments.

\clearpage

\subsection{Phase transition model}\label{CsPbBr3_PT}
To avoid the influence of volume expansion on diffusion, all equilibration and production runs are performed at constant volume. The cell volumes at different temperatures are deduced from constant temperature MD runs using a \ce{CsPbBr3} phase transition model.  The model is trained using \supercell{2}{2}{2} (8 units of \ce{CsPbI_{3}}) pseudo-cubic cells in all three perovskite phases of \ce{CsPbBr_{3}}. We start the training in the cubic phase at 500 K, followed by the tetragonal phase at 370 K, and finally in the orthorhombic phase at 370 K and 150 K. All these runs are performed at $\mathrm{10^{5}}$ Pa pressure for 100 ps with MD time-steps of 10 fs. For the DFT calculations, we use the same functional, PAW potentials, and kinetic energy cutoff as in Section~\ref{CsPbBr3_training}, energy convergence criteria of $\mathrm{10^{-4}}\,\mathrm{eV}$, and a $\mathrm{2\times2\times2}$ Monkhorst-Pack $k$-point grid. To construct the MLFF, cutoff radii of $6\,\text{\AA}$ and $5\,\text{\AA}$ are used for the radial and the angular descriptors, respectively. The atomic positions are expanded using Gaussian distributions of width $0.5\,\text{\AA}$. The radial descriptor is expanded using 6 radial basis functions, and the angular descriptor is expanded using 6 radial basis functions and spherical harmonics with a maximum angular momentum quantum number of 9. Gaussian kernels of polynomial power 4 are used, where the radial and angular descriptors have weights 0.7 and 0.3, respectively.

\begin{figure}
    \centering
    \includegraphics{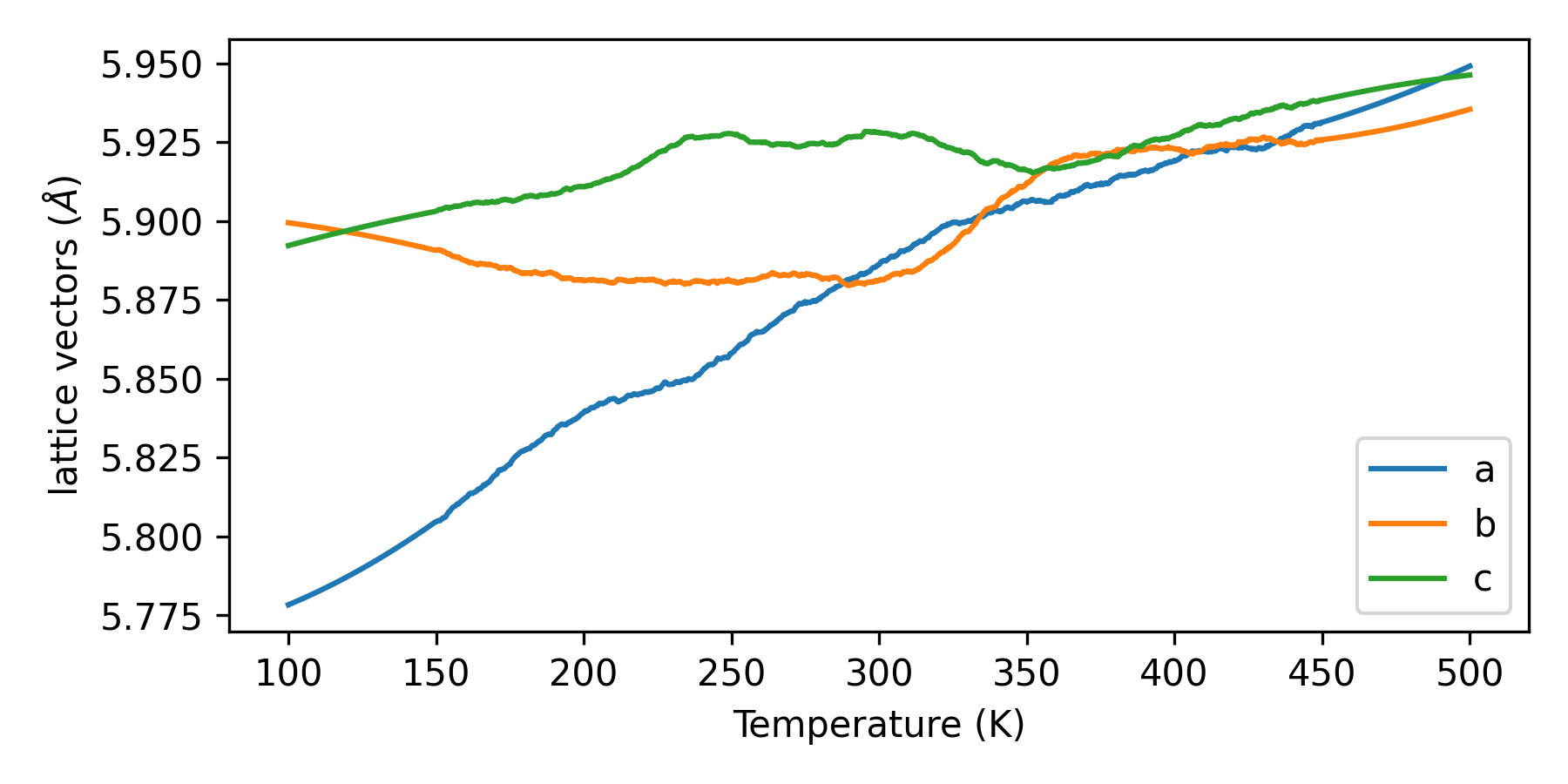}
    \caption{Pseudo-cubic unit cell lattice vectors of \ce{CsPbBr_{3}} as function of temperature in the heating MD run.}
    \label{fig:CsPbBr3-phase-transition}
\end{figure}

The accuracy of this model is validated by performing a heating MD run from 100 K to 500 K for \SI{0.8}{ns} with an MD time step of \SI{10}{fs} using a \supercell{3}{3}{2} (72 units of \ce{CsPbI_{3}}) orthorhombic supercell. The window averaged pseudo-cubic lattice vectors with the window length \SI{0.16}{ns} from this run are given in Figure~\ref{fig:CsPbBr3-phase-transition}, and the overall phase diagram closely resembles that observed in experiments \cite{nasstrom2020dependence}.

Using this force field, we perform \SI{100}{ps} long constant temperature MD runs using \supercell{6}{6}{6} cubic (216 units of \ce{CsPbBr_{3}}) supercells in the \textit{NpT} ensemble at various temperatures between \SI{400}{K} and \SI{500}{K}. The variation in the cube root of the final volume with temperature from these runs is extrapolated to get lattice constants at various temperatures between \SI{500}{K} and \SI{600}{K}.

\clearpage

\subsection{Production runs and diffusion coefficients}\label{CsPbBr_production}
Following training and validation, the force fields are used to perform multiple MD runs on \supercell{6}{6}{6} cubic supercells (216 units of \ce{CsPbBr3}) with one bromide point defect at temperatures between \SI{500}{K} and \SI{600}{K}. The volume is kept constant during each run with lattice parameters at different temperatures extracted from the constant temperature MD runs on pristine \ce{CsPbBr3}. The lattice constants at different temperatures are given in Table~\ref{tab:table4}.

\begin{table}[]
    \centering
    \begin{tabular}{|c c|}
        \hline
        Temperature (K) & Lattice constants ($\text{\AA}$) \\
        \hline
        500 & 5.939 \\
        525 & 5.943 \\
        550 & 5.947 \\
        575 & 5.950 \\
        600 & 5.954 \\
        \hline
    \end{tabular}
    \caption{\textmd{Lattice constants of \ce{CsPbBr_{3}} unit cell at different temperatures in cubic phase.}}
    \label{tab:table4}
\end{table}

The structures are first equilibrated at their target temperature for \SI{100}{ps} with timesteps of \SI{2}{fs} in an \textit{NVT} ensemble using Langevin thermostat with friction coefficients set to \SI{3}{ps^{-1}} for all atomic species. The atomic positions and velocities of the final frames from the equilibration runs are used as the starting point for \SI{2}{ns} long production runs in an \textit{NVT} ensemble using Nose-Hoover thermostat with Nose mass of \SI{0.08}{ps}. To ensure the proper sampling of diffusion coefficients, we perform at least 5 production runs at each temperature. 

\begin{figure}
    \centering
    \includegraphics{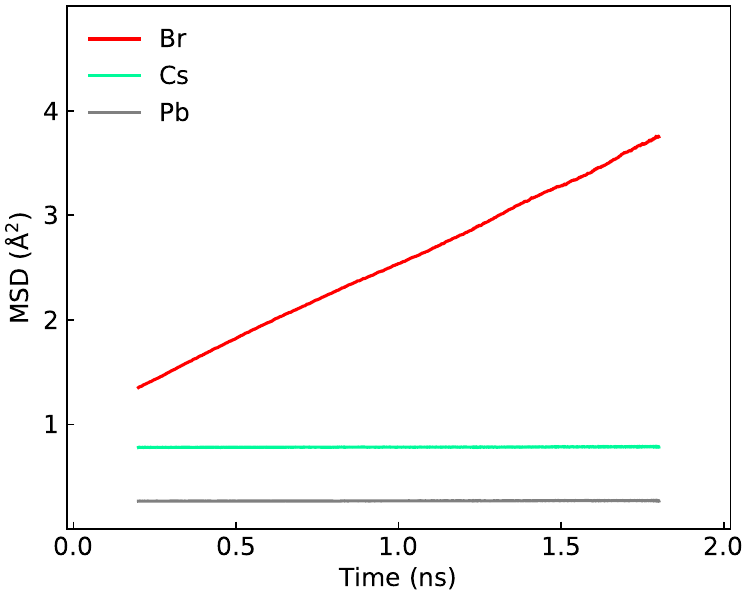}
    \caption{Mean Squared displacement (MSD) curves decomposed to the atomic species over simulation time for \chargeddefect{I}{-} defect in \ce{CsPbBr3} at \SI{550}{K}.}
    \label{fig:CsPbBr3_MSD}
\end{figure}

Multiple migration events are observed for all defects, except \chargeddefect{V}{-}. To quantify the migration behavior of the defects, the mean squared displacement (MSD) was plotted over time for each atomic species using the MDAnalysis Python library \cite{michaud2011mdanalysis,Maginn2018best}; one such plot is given in Figure~\ref{fig:CsPbBr3_MSD}. The MSD is calculated using
\begin{equation}
    \text{MSD}(r_{d}) = \Bigl \langle \frac{1}{N}\sum_{i=1}^{N}|r_{d}-r_{d}(t_{0})|^{2}\Bigr \rangle_{t_{0}},
    \label{eqn:equation_MSD}
\end{equation}
where $N$ is the number of atoms of a particular atomic species, and $r_{d}$ are their coordinates in $d$ dimensions (3 for our systems). The Diffusion coefficient $D$ is calculated using
\begin{equation}
    D = \frac{N}{2d}\lim_{t\rightarrow\infty}\frac{d}{dt}\text{MSD}(r_{d}),
    \label{eqn:equation_DC}
\end{equation}
which is proportional to the slope of the MSD curve, where the factor $N$ ensures that the diffusion coefficient is defect concentration independent.

\begin{figure}
    \centering
    \includegraphics{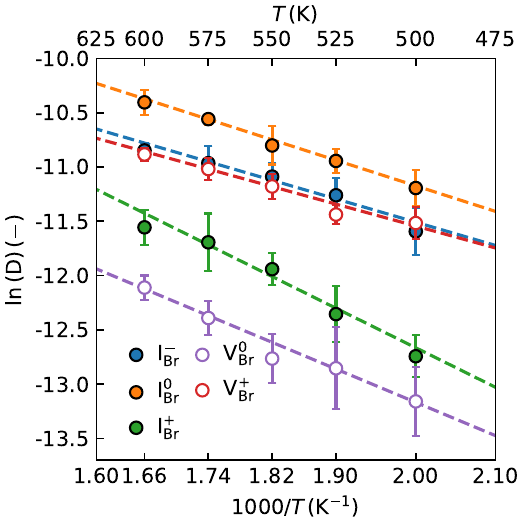}
    \caption{Temperature-dependent diffusion coefficients of halide point defects obtained from the MD simulations with the MLFFs for halide point defects in \ce{CsPbBr3}. The filled symbols represent halide interstitials, and the open symbols represent halide vacancies. The dashed lines represent the fits to an Arrhenius expression, and the error bars
represent the standard error in mean at each point.}
    \label{fig:CsPbBr3_diffusion_curve}
\end{figure}

The diffusion behavior for all five mobile species can be fitted by an Arrhenius relation
\begin{equation}
    D = D_{0}\exp\left(-\frac{E_\mathrm{a}}{k_\mathrm{B}T}\right)
    \label{eqn:equationAR}
\end{equation}
where $k_\mathrm{B}$ is the Boltzmann constant, $E_\mathrm{a}$ the activation energy, and $D_{0}$ the pre-exponental factor. The fits are given in Figure~\ref{fig:CsPbBr3_diffusion_curve}.

The $E_\mathrm{a}$, $D_{0}$, and diffusion coefficients at room temperature ($D_{\mathrm{300K}}$), extrapolated from the Arrhenius relation (Eq.~\ref{eqn:equationAR}), are given in Table~\ref{tab:diffusion-data-CsPbBr3}. At this temperature, interstitials and vacancies in their most stable charge states, \chargeddefect{I}{-} and \chargeddefect{V}{+}, have similar diffusion coefficients (7.09 vs 7.38 $\mathrm{\times10^{-7}}$ \SI{}{cm^{2}s^{-1}}). The neutral defects, with \chargeddefect{I}{-} capturing a hole or \chargeddefect{V}{+} capturing an electron, behave differently. The bromide interstitial \chargeddefect{I}{0} migrates at a similar rate to \chargeddefect{I}{-}, while the vacancy \chargeddefect{V}{0} migrates at an order of magnitude slower than \chargeddefect{V}{+}. Finally, \chargeddefect{I}{-} capturing two holes makes the interstitial \chargeddefect{I}{+} much less mobile, whereas \chargeddefect{V}{+} capturing two electrons makes the vacancy \chargeddefect{V}{-} immobile in the present simulations.

\begin{table}
    \centering
    \begin{tabular}{|c c c c|}
     \hline
     System & ${E_\mathrm{a}\,\,\mathrm{(eV)}}$ & ${D_{0}\,\,(\times10^{-3}\,\mathrm{cm}^{2}\mathrm{s}^{-1})}$ & ${D_{300K}\,\,(\times10^{-7}\,\mathrm{cm}^{2}\mathrm{s}^{-1})}$\\
     \hline
     $\mathrm{I_{Br}^{-}}$ & {$\mathrm{0.19\pm0.03}$} & {$\mathrm{0.75\pm0.43}$} & {7.09}\\
     $\mathrm{I_{Br}^{0}}$ &
     {$\mathrm{0.20\pm0.02}$} & {$\mathrm{1.54\pm0.74}$} &
     {6.72}\\
     $\mathrm{I_{Br}^{+}}$ &
     {$\mathrm{0.32\pm0.04}$} & {$\mathrm{4.76\pm3.75}$} &
     {0.20}\\
     $\mathrm{V_{Br}^{0}}$ &
     {$\mathrm{0.26\pm0.04}$} & {$\mathrm{0.84\pm0.81}$} &
     {0.36}\\
     $\mathrm{V_{Br}^{+}}$ &
     {$\mathrm{0.17\pm0.02}$} & {$\mathrm{0.53\pm0.22}$} &
     {7.38}\\
     \hline
    \end{tabular}
    \caption{\textmd{Activation energies (${E_\mathrm{a}}$) and pre-exponential factors (${D_{0}}$) extracted from the Arrhenius fits, and extrapolated diffusion constants (${D_{300\mathrm{K}}}$) at room temperature.}}
    \label{tab:diffusion-data-CsPbBr3}
\end{table}

\clearpage

\section{Model efficiency with chemical species}\label{model_efficiency}
\begin{figure}
    \centering
    \includegraphics[]{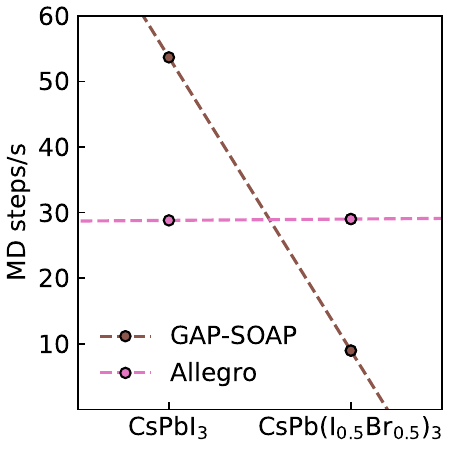}
    \caption{Number of MD steps per second of simulation time for the \ce{CsPbI3} and \mixedsystem{0.5}{0.5} phase transition models in the GAP-SOAP and Allegro architecture.}
    \label{fig:efficiency_loss}
\end{figure}

In addition, we compare the efficiency of the Allegro model with the GAP-SOAP model as implemented in VASP, these comparisons are given in Figure ~\ref{fig:efficiency_loss}. Here, the MD runs using the GAP-SOAP model are performed in VASP using 8 cores on AMD EPYC 9654P CPUs, and MD using the Allegro model is performed in LAMMPS using a single NVIDIA Tesla v100 GPU.

For the GAP-SOAP model, the efficiency drops significantly from \SI{53.66}{steps/s} to \SI{8.96}{steps/s} when the number of chemical species in the system is increased from three (\ce{CsPbI3}) to four (\mixedsystem{0.5}{0.5}) (Figure ~\ref{fig:efficiency_loss}). On the other hand, the efficiency of the Allegro model does not scale with the number of chemical species in the system, staying around \SI{29}{steps/s} for both \ce{CsPbI3} and \mixedsystem{0.5}{0.5} (Figure ~\ref{fig:efficiency_loss}).

\clearpage

\section{\mixedsystem{x}{1-x} defect migration model training}\label{mixed_training}
Training structures are sampled from three constant temperature MD runs using the on-the-fly learning procedure as implemented in VASP, for both negatively charged halide interstitials and positively charged halide vacancies. These runs are performed for \SI{100}{ps} with an MD timestep of \SI{2}{fs}, using \supercell{2}{2}{2} cubic supercells of \mixedsystem{0.5}{0.5} (8 units) with one halide point defect in them. These runs are performed at \SI{700}{K}, \SI{600}{K}, and \SI{750}{K} in \textit{NpT} ensembles at $\mathrm{10^{5}}$ Pa pressure. To maintain constant temperature and pressure, Parinello-Rahman dynamics is used with friction coefficients set to \SI{3}{ps^{-1}} for all atomic species and lattice degrees of freedom.

A total of 4231 and 4740 structures are sampled for the halide interstitial and halide vacancy, respectively. We run DFT single point calculations on all these structures using the same functional, PAW potentials for \ce{Br},\ce{Cs}, and \ce{Pb}, and kinetic energy cutoff as in Section~\ref{CsPbBr3_training} of SI note~\ref{CsPbBr3}, ($\mathrm{5s^{2}5p^{5}}$) PAW potential for \ce{I}, energy convergence criteria of $\mathrm{10^{-6}}$ \SI{}{eV}, and \supercell{2}{2}{2} Monkhorst-Pack \textit{k}-grid to generate the final training and validation sets.

The Allegro model \cite{Batzner2022equivariant,Musaelian2023learning} is trained using a radial cutoff of \SI{6.5}{\AA} with interatomic distances projected onto a radial basis using trainable Bessel functions, and 2 tensor product layers; 32 ordered-pair tensor features are used, expanded using spherical harmonics with a maximum angular quantum number ($\mathrm{L_{max}}$) of 2, while preserving full \textit{O}(3) symmetry. The 2-body latent multilayer perceptron (MLP) had the dimensions [128, 256, 512, 1024] and SiLU nonlinearity, and the latent MLP had the dimensions [1024, 1024, 1024] also with SiLU nonlinearity. To predict the pair energies, the final MLP has dimensions [128] without nonlinearity.

The training and validation sets consist of 7177 and 1794 structures, respectively, a 80\% and 20\% split of the total number of structures. These structures are shuffled at each epoch during training. The training is performed using the total energy of the system, forces acting on the atoms, and stress tensors. We use the per-atom MSE loss function, with the weight set to 1 for both energy and forces. The Adam optimizer in PyTorch is used with the default parameters $\beta_{1}=0.9$, $\beta_{2}=0.99$, and $\epsilon=10^{-8}$. A learning rate of 0.001 and a batch size of 5 are used, and training ran for 592 epochs.

\begin{figure}
    \centering
    \includegraphics[]{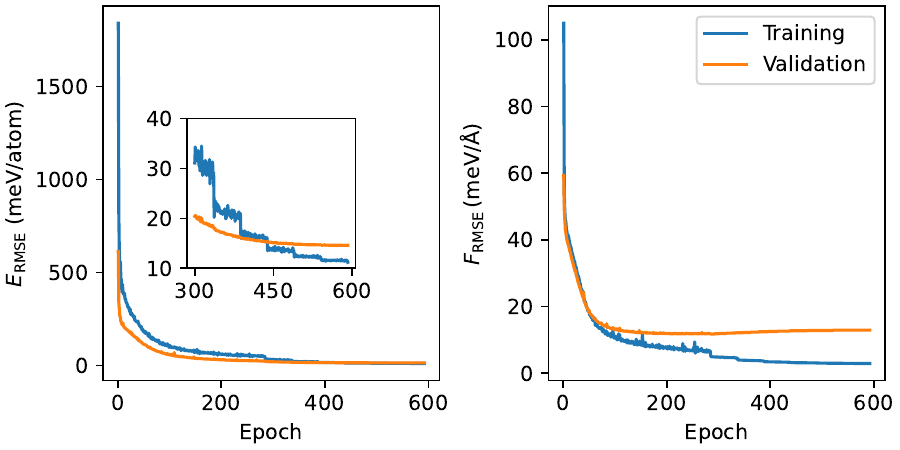}
    \caption{Root mean squared errors in energy ($E_{\mathrm{RMSE}}$) and forces ($F_{\mathrm{RMSE}})$ during model training and model validation of the \mixedsystem{x}{1-x} defect migration model.}
    \label{fig:mixed__defect_model_erros}
\end{figure}

The RMSE in energy and forces during model training and model validation are given in Figure~\ref{fig:mixed__defect_model_erros}. The final model has the training errors of \SI{11.2}{meV/atom} in energy and \SI{2.88}{meV/\AA} in forces, and validation errors of \SI{14.6}{meV/atom} in energy and \SI{12.90}{meV/\AA} in forces.

\clearpage

\section{\mixedsystem{x}{1-x} defect migration model validation}\label{mixed_validation}
\begin{figure}
    \centering
    \includegraphics[]{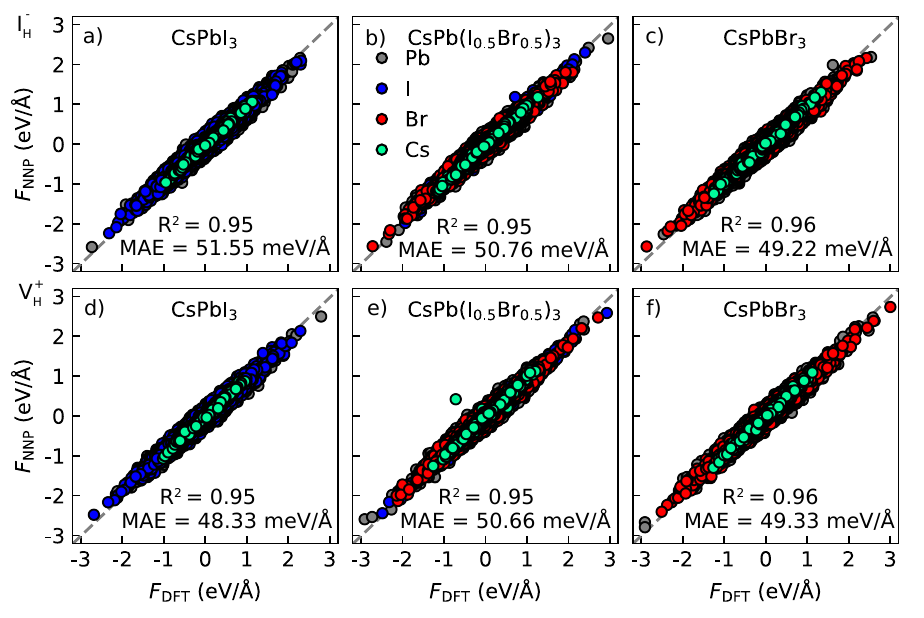}
    \caption{Comparison between forces acting on all atoms calculated by the NNP with forces calculated using DFT for \supercell{6}{6}{6} supercells with one halide interstitial (a-c), or one halide vacancy (d-f) for \ce{CsPbI3}, \mixedsystem{0.5}{0.5}, and \ce{CsPbBr3}, along with corresponding $\mathrm{R^{2}}$ values and mean absolute errors (MAE).}
    \label{fig:NNP_force_comparison_all}
\end{figure}

To check the accuracy of the model, we sample 20 structures from \SI{0.5}{ns} MD runs at \SI{600}{K} performed using \supercell{6}{6}{6} supercells with one halide point defect for pure \ce{CsPbI3}, { \mixedsystem{0.5}{0.5} with \ce{I} and \ce{Br} positioned randomly on halide sites}, and pure \ce{CsPbBr3}, and compare forces calculated using DFT with forces calculated using the NNP for these structures. The forces categorized according to the atomic species acting on all atoms, with the corresponding $\mathrm{R^{2}}$ values and mean absolute errors (MAE) are given in Figure~\ref{fig:NNP_force_comparison_all}. From these comparisons, we note that the $\mathrm{R^{2}}\,\geq\,0.95$ and $\mathrm{MAE\,\leq\,\SI{51.55}{meV/\angstrom}}$ for all systems. Considering that the forces acting on these atoms are of the order of \SI{1}{eV/\angstrom}, the NNP is highly accurate in calculating forces acting on all atoms. On top of that, the NNP is also highly transferable to different halide compositions, with the values of MAE being very close for all systems, between \SI{48.33}{meV/\angstrom} and \SI{51.55}{meV/\angstrom}.

\begin{figure}
    \centering
    \includegraphics[]{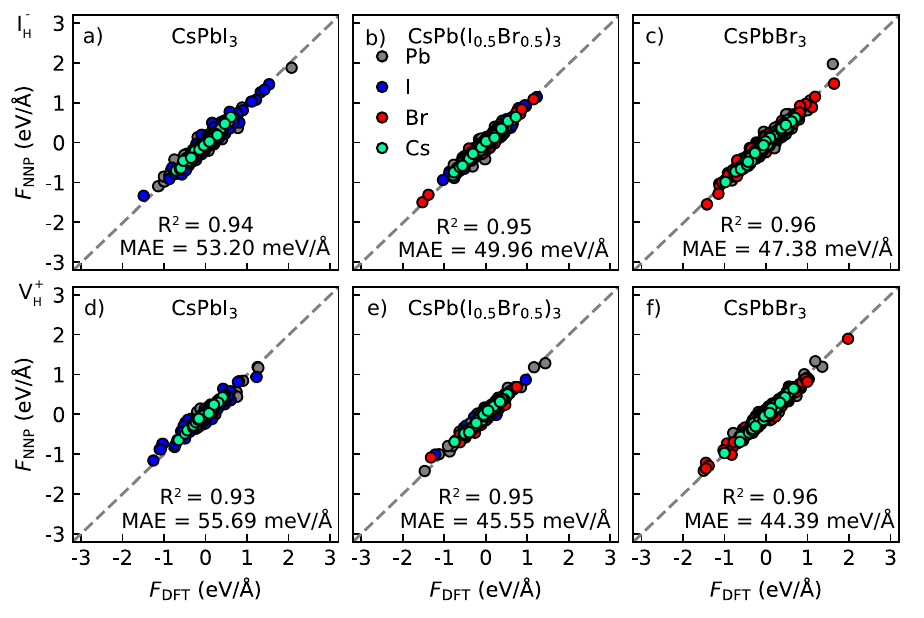}
    \caption{Comparison between forces acting on atoms in the defect environment calculated by the NNP with forces calculated using DFT for \supercell{6}{6}{6} supercells with one halide interstitial (a-c), or one halide vacancy (d-f) for \ce{CsPbI3}, \mixedsystem{0.5}{0.5}, and \ce{CsPbBr3}, along with corresponding $\mathrm{R^{2}}$ values and mean absolute errors (MAE).}
    \label{fig:NNP_force_comparsion_defect}
\end{figure}

We also compare the forces acting on atoms close to the defect environment, these comparisons are given in Figure~\ref{fig:NNP_force_comparsion_defect}. From these comparisons, we conclude that the NNP also calculates forces acting on atoms in the defect environment with high accuracy, with the $\mathrm{R^{2}}\,\geq\,0.94$ and $\mathrm{MAE\,\leq\,\SI{55.69}{meV/\angstrom}}$ for all systems.

Along with the forces, we also compare the energies predicted by the NNP with those calculated using DFT along defect migration paths. We take 4 different migration paths, one for each halide species, for both vacancy and interstitial. These paths are constructed using climbing image nudged elastic band (CI-NEB) calculations on \supercell{2}{2}{1} orthorhombic supercells of { \mixedsystem{x}{1-x} (16 units) with \ce{I} and \ce{Br} placed randomly on halide sites} with one halide point defect for all systems except interstitial migrating through Br ($\mathrm{I_{X}^{-}}$ (\ce{Br})), for which convergence is not achieved using a orthorhombic supercell, hence a \supercell{2}{2}{2} cubic supercell (8 units of \mixedsystem{x}{1-x}) is used. 3 intermediate images, connected by springs with spring constant $\mathrm{5\,eV/\text{\AA}^{2}}$, are optimized, using a plane wave cut-off energy of \SI{300}{eV}, and \supercell{2}{2}{3} and \supercell{4}{4}{4} Monkhorst-Pack $k$-point grid for the orthorhombic and cubic supercells, respectively. The computed migration barriers are given in Table ~\ref{tab:NNP_NEB_barriers}. The difference between the migration barriers calculated using DFT and the NNP is less than \SI{0.08}{eV} for all systems, indicating the high accuracy and transferability of the NNP in calculating the migration barriers.

\begin{table}[]
    \centering
    \begin{tabular}{|c c c|}
        \hline
        System & $E\mathrm{_b^{DFT}}$ (\SI{}{eV}) & $E\mathrm{_b^{NNP}}$ (\SI{}{eV}) \\
        \hline
        $\mathrm{I_{X}^{-}}$ (\ce{Br}) & 0.429 & 0.358 \\
        $\mathrm{I_{X}^{-}}$ (\ce{I}) & 0.186 & 0.113 \\
        $\mathrm{V_{X}^{+}}$ (\ce{Br}) & 0.362 & 0.389 \\
        $\mathrm{V_{X}^{+}}$ (\ce{I}) & 0.294 & 0.277 \\
        \hline
    \end{tabular}
    \caption{\textmd{Defect migration barriers ($E\mathrm{_{b}}$) calculated using DFT and the NNP for CI-NEB migration paths.}}
    \label{tab:NNP_NEB_barriers}
\end{table}

{ Finally, the NNP optimized defect structures are compared with those optimized using DFT to check the accuracy of the model in capturing subtle structural differences between defect configurations. \supercell{2}{2}{1} orthorhombic supercells of \mixedsystem{0.5}{0.5} (16 units) with \ce{I} and \ce{Br} positioned randomly on halide sites, with one halide point defect, interstitial, in either one of the three configurations, or vacancy, are used. The DFT optimization is performed using the same functional, PAW potentials, and kinetic energy cutoff as in Section~\ref{mixed_training}, energy convergence criteria of $\mathrm{10^{-5}}$ \SI{}{eV} between electronic steps, force convergence criteria of $\mathrm{10^{-2}}$ \SI{}{eV/\angstrom} between ionic steps, and \supercell{4}{4}{6} Monkhorst-Pack \textit{k}-grid. The NNP optimization is performed using force convergence criteria of $\mathrm{10^{-2}}$ \SI{}{eV/\angstrom}. The average \ce{Pb-I} and \ce{Pb-Br} bond distances are given in ~\ref{tab:average_distances}, the NNP calculated distances are in excellent agreement with the DFT calculated ones.

\begin{table}[]
    \centering
    \begin{tabular}{|c c c c c|}
        \hline
        { Defect} & \multicolumn{2}{c}{ $\mathrm{\overline{r}_{Pb-I}\,(\AA)}$} & \multicolumn{2}{c|}{ $\mathrm{\overline{r}_{Pb-Br}\,(\AA)}$} \\
        & { DFT} & { NNP} & { DFT} & { NNP} \\
        \hline
        { $\mathrm{I_{I-I}^{-}}$} & { 3.20} & { 3.20} & { 3.02} & { 3.01} \\
        { $\mathrm{I_{I-Br}^{-}}$} & { 3.21} & { 3.20} & { 3.02} & { 3.01} \\
        { $\mathrm{I_{Br-Br}^{-}}$} & { 3.22} & { 3.22} & { 3.00} & { 3.00}\\
        { $\mathrm{V_{X}^{+}}$} & { 3.19} & { 3.19} & { 3.00} & { 3.00} \\
        \hline
    \end{tabular}
    \caption{\textmd{{ Average \ce{Pb-I} ($\mathrm{\overline{r}_{Pb-I}}$) and \ce{Pb-Br} ($\mathrm{\overline{r}_{Pb-Br}}$) bond distance in \mixedsystem{0.5}{0.5} structures with one halide point defect, interstitial, in either one of the three configurations, or vacancy, optimized using DFT and the NNP. }}}
    \label{tab:average_distances}
\end{table}

Further, the distance between the halide atoms forming the interstitial, and the \ce{Pb} atoms adjacent to the vacancy are given in table~\ref{tab:defect_distances}. Distances for all the NNP optimized structures are also in very good agreement with the DFT optimized structures.

\begin{table}[]
    \centering
    \begin{tabular}{|c c c|}
        \hline
        { Defect} & { $\mathrm{r_{DFT}\,(\AA)}$} & { $\mathrm{r_{NNP}\,(\AA)}$} \\
        \hline
        { $\mathrm{I_{I-I}^{-}}$} & { 3.85} & { 3.79} \\
        { $\mathrm{I_{I-Br}^{-}}$} & { 3.65} & { 3.65} \\
        { $\mathrm{I_{Br-Br}^{-}}$} & { 3.57} & { 3.56} \\
        { $\mathrm{V_{X}^{+}}$} & { 6.50} & { 6.50} \\
        \hline
    \end{tabular}
    \caption{\textmd{{ Distance between halide atoms forming the interstitial in the \ce{Pb-II-Pb} ($\mathrm{I_{I-I}^{-}}$), \ce{Pb-IBr-Pb} ($\mathrm{I_{I-Br}^{-}}$), and the \ce{Pb-BrBr-Pb} ($\mathrm{I_{Br-Br}^{-}}$) configurations, and the distance between the \ce{Pb} atoms adjacent to the halide vacancy ($\mathrm{V_{X}^{+}}$) optimized using DFT and the NNP.}}}
    \label{tab:defect_distances}
\end{table}

}

\clearpage

\section{\mixedsystem{x}{1-x} phase transition}\label{mixed_PT}
Training structures are sampled from six constant temperature MD runs using the on-the-fly learning procedure as implemented in VASP. These runs are performed for \SI{50}{ps} with timesteps of \SI{10}{fs}, using \supercell{2}{2}{2} pseudo-cubic supercells of \mixedsystem{0.5}{0.5} (8 units) in all three perovskite phases. The starting structures are constructed such that \ce{Pb} atoms are coordinated to different numbers of \ce{I} and \ce{Br} atoms. This ensures that diverse halide environments are captured for \ce{Pb} atoms. We start training in the cubic phase at \SI{700}{K} and \SI{500}{K}, following that we train in the tetragonal phase at \SI{500}{K} and \SI{370}{K}, and finally in the orthorhombic phase at \SI{350}{K} and \SI{150}{K}. All these runs are performed in an \textit{NpT} ensemble with Parinello-Rahman dynamics with friction coefficients set to \SI{3}{ps^{-1}} for all atomic species and lattice degrees of freedom.

A total of 1062 structures are sampled, and we run DFT single-point calculations on all these structures using the same DFT parameters as in SI note~\ref{mixed_training} to generate the final training and validation sets. 

\begin{figure}
    \centering
    \includegraphics[]{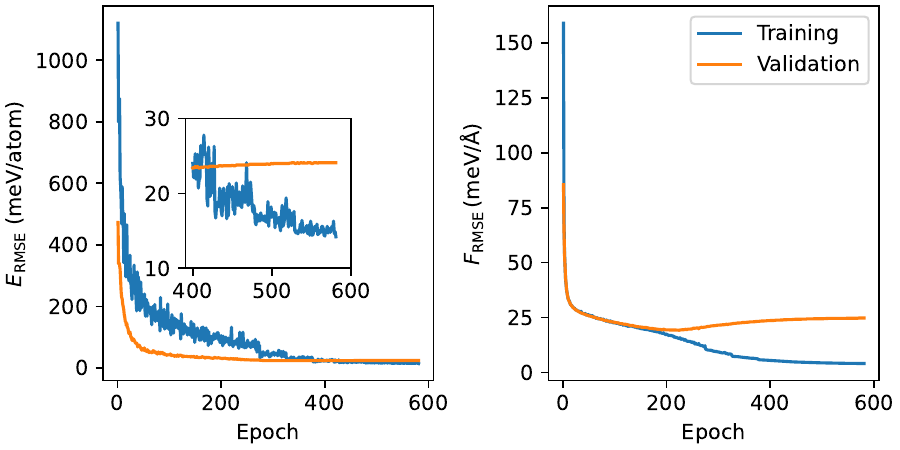}
    \caption{Root mean squared errors in energy ($E_{\mathrm{RMSE}}$) and forces ($F_{\mathrm{RMSE}})$ during model training and model validation of the \mixedsystem{x}{1-x} phase transition model.}
    \label{fig:mixed_PT_model_errors}
\end{figure}

An Allegro model is trained using the same model parameters as in SI note 3, with the training and validation sets consisting of 850 and 212 structures, respectively, and the training ran for 581 epochs. The RMSE in energy and forces during model training and model validation are given in Figure ~\ref{fig:mixed_PT_model_errors}. The final model has the training errors of \SI{14.2}{meV/atom} in energy and \SI{4.08}{meV/\angstrom} in forces, and validation errors of \SI{24.1}{meV/atom} in energy and \SI{24.8}{meV/\angstrom} in forces.

\clearpage

This NNP is used to perform constant temperature MD runs in an \textit{NpT} ensemble to obtain equilibrated volumes at different temperatures and different halide mixing ratios for \mixedsystem{x}{1-x}.

\begin{figure}
    \centering
    \includegraphics[]{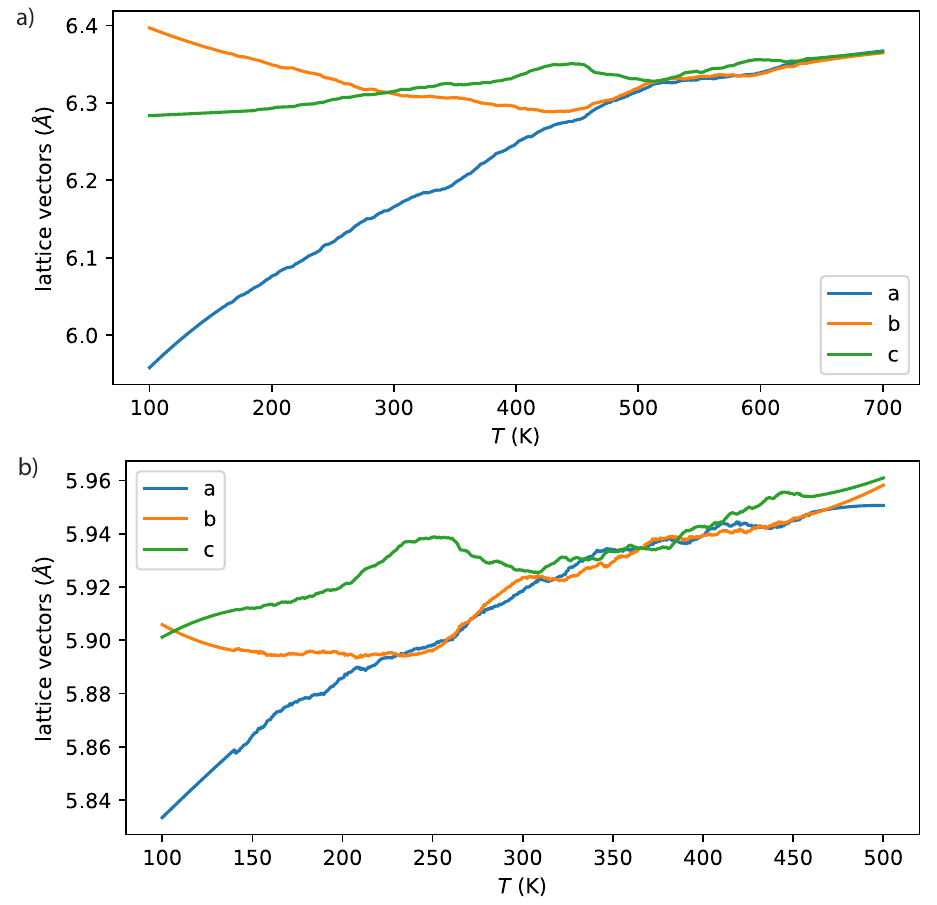}
    \caption{Pseudo-cubic unit cell lattice vectors of a) \ce{CsPbI_{3}} and b) \ce{CsPbBr3} as functions of temperature in the heating MD runs performed using the \mixedsystem{x}{1-x} phase transition NNP.}
    \label{fig:CsPbIBr_PT_heating}
\end{figure}

The accuracy and transferability of this model are validated by performing heating runs from \SI{100}{K} to \SI{500}{K} for \SI{0.4}{ns} using a time step of \SI{2}{fs} for \ce{CsPbBr3}, and from \SI{100}{K} to \SI{700}{K} for \SI{0.6}{ns} using a time step of \SI{2}{fs} for \ce{CsPbI3}. These runs are performed using \supercell{3}{3}{2} orthorhombic supercells (72 units of \ce{CsPbX3}). The window-averaged pseudo-cubic lattice vectors with the window lengths \SI{0.12}{ns} for \ce{CsPbI3} and \SI{0.08}{ns} for \ce{CsPbBr3} from these runs are given in Figure ~\ref{fig:CsPbIBr_PT_heating}a and b, respectively. The overall phase diagram closely resembles that observed in experiments \cite{marronnier2018anharmonicity,nasstrom2020dependence}.

\clearpage

\section{Halide diffusion in \mixedsystem{0.5}{0.5}}\label{mixed_migration}
\SI{2}{ns} long MD runs are performed at different temperatures between \SI{500}{K} and \SI{600}{K} using \supercell{6}{6}{6} supercells (216 units of \mixedsystem{0.5}{0.5}) with one halide point defect in LAMMPS \cite{thompson2022lammps}. The lattice constants are obtained using the procedure described in SI note~\ref{mixed_PT}, and are given in Table ~\ref{tab:CsPb(I0.5Br0.5)3_lattice_constants}.

\begin{table}[]
    \centering
    \begin{tabular}{|c c|}
        \hline
        Temperature (K) & Lattice constants ($\text{\AA}$) \\
        \hline
        500 & 6.143 \\
        525 & 6.148 \\
        550 & 6.154 \\
        575 & 6.159 \\
        600 & 6.164 \\
        \hline
    \end{tabular}
    \caption{\textmd{Lattice constants of \mixedsystem{0.5}{0.5} unit cell at different temperatures in cubic phase.}}
    \label{tab:CsPb(I0.5Br0.5)3_lattice_constants}
\end{table}

\begin{figure}
    \centering
    \includegraphics[]{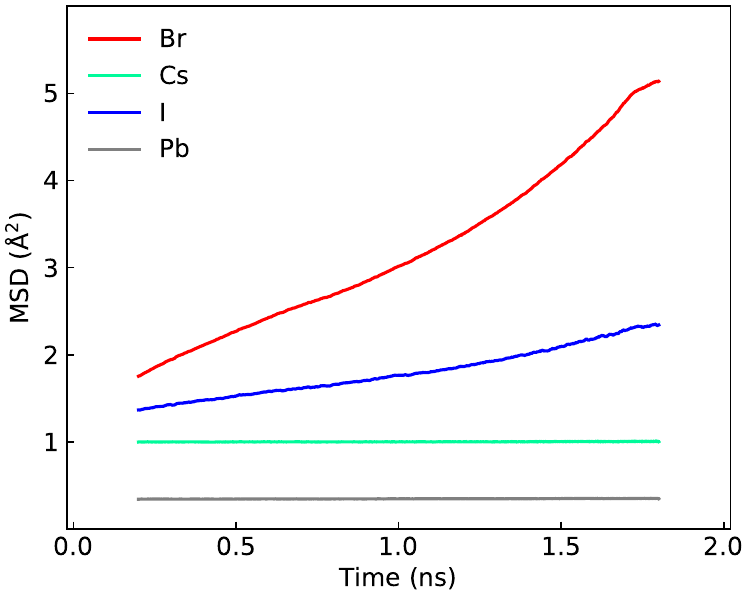}
    \caption{Mean squared displacement (MSD) curves decomposed to the atomic species over simulation time for halide interstitial in \mixedsystem{0.5}{0.5} at \SI{550}{K}.}
    \label{fig:msd_CsPbIBr}
\end{figure}

The structures are first equilibrated to the target temperature for \SI{100}{ps} with timesteps of \SI{2}{fs} in an \textit{NVT} ensemble using a Langevin thermostat with the relaxation time set to \SI{0.33}{ps}. Following equilibration, \SI{2}{ns} long production runs with timesteps of \SI{2}{fs} are performed in an \textit{NVT} ensemble using a Nose-Hoover thermostat with the relaxation time set to \SI{0.2}{ps}. To ensure proper sampling of diffusion coefficients, 5 runs are performed at each temperature. The MSD is plotted over time for each atomic species, one such plot is given in Figure ~\ref{fig:msd_CsPbIBr}. The diffusion coefficients for halide species are calculated using the procedure mentioned in Section~\ref{CsPbBr_production} for the SI note~\ref{CsPbBr3}, and are given in Figure 2 in the main text.

\clearpage

\section{Halide bridge configuration}\label{mixed_bridge}
\begin{figure}
    \centering
    \includegraphics[]{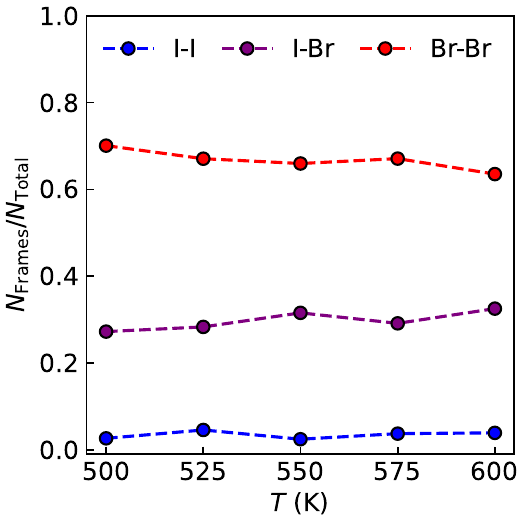}
    \caption{Population ($N_{\mathrm{Frames}}/N_{\mathrm{Total}}$) of all three interstitial bridge configurations at different temperatures.}
    \label{fig:interstitial_bridge_population}
\end{figure}

To check which halide bridge configuration is the most energetically favorable, we do a population analysis of the three different configurations from the MD simulations, which are given in Figure ~\ref{fig:interstitial_bridge_population}. As evident from the figure, the \ce{Pb-BrBr-Pb} is the most observed configuration, followed by the \ce{Pb-IBr-Pb} configuration, both of which are observed significantly more than the \ce{Pb-II-Pb} configuration at all simulated temperatures.

\clearpage

\section{Halide fluctuations}\label{mixed_fluctuations}
\begin{figure}
    \centering
    \includegraphics[]{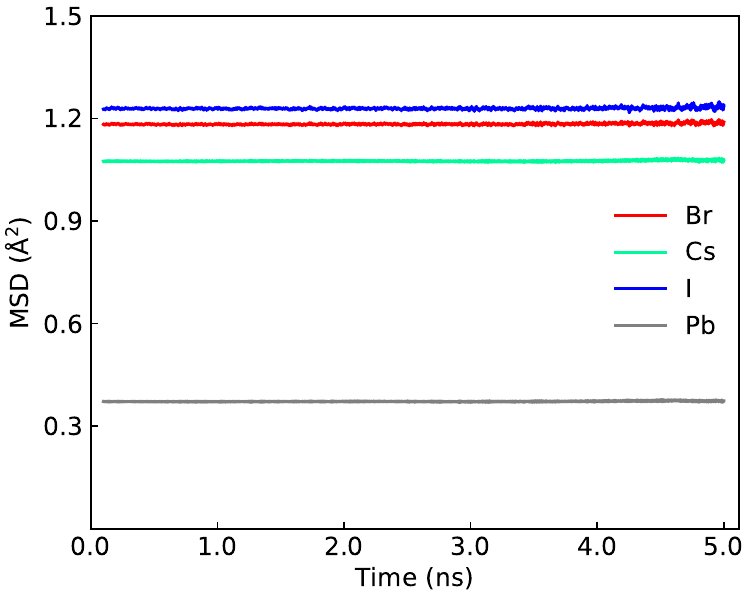}
    \caption{Mean squared displacement (MSD) curves decomposed to the atomic species over simulation time for pristine \mixedsystem{0.5}{0.5} at \SI{600}{K}.}
    \label{fig:msd_CsPbIBr_pristine}
\end{figure}

To check which halide species fluctuates more around their mean position, we perform a \SI{5}{ns} long MD run using a pristine \supercell{6}{6}{6} cubic supercell of \mixedsystem{0.5}{0.5} at \SI{600}{K}. The structure is first equilibrated to the target temperature for \SI{100}{ps}, and both equilibration and production runs are performed in an \textit{NVT} ensemble using the same thermostat settings as described in SI note~\ref{mixed_migration}. The MSD is plotted over time for each atomic species, and the plot is given in Figure ~\ref{fig:msd_CsPbIBr_pristine}. The MSD value is \SI{1.23}{\angstrom^{2}} for \ce{I} and \SI{1.18}{\angstrom^{2}} for \ce{Br}, indicating that \ce{I} fluctuate more around their mean positions than \ce{Br}.

\clearpage

\section{Bulk interfaces}\label{bulk_interfaces}
\begin{figure}
    \centering
    \includegraphics[]{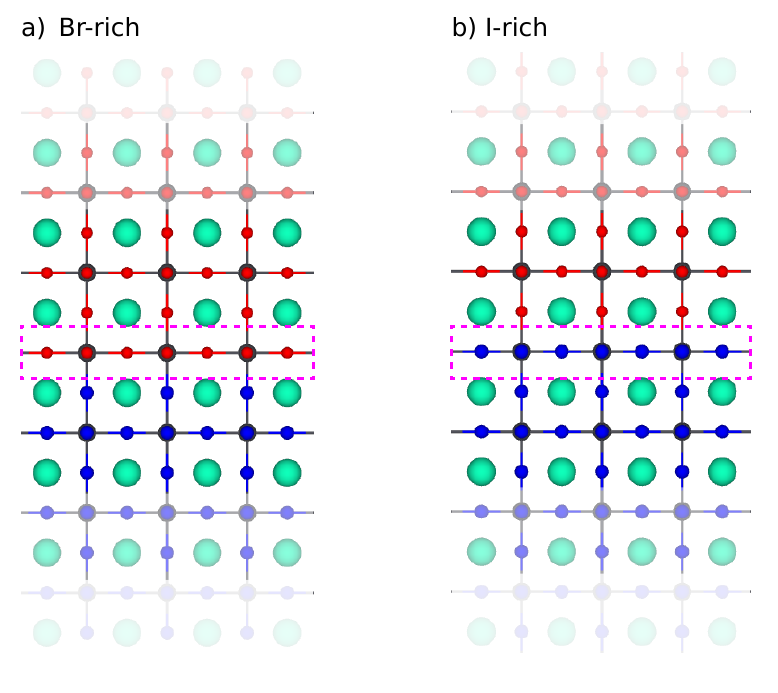}
    \caption{a) \ce{Br}-rich and b) \ce{I}-rich bulk interfaces, the interface layer is surrounded by dashed pink lines.}
    \label{fig:bulk_interfaces_definition}
\end{figure}

To study the influence of halide migration on halide phase mixing in \mixedsystem{x}{1-x}, migration behavior of halide point defects along bulk interfaces is studied. We define two bulk interfaces, the \ce{Br}-rich interface, where the \ce{Pb} atoms on the interface are coordinated to five \ce{Br} and one \ce{I} atoms (Figure ~\ref{fig:bulk_interfaces_definition}a), and the \ce{I}-rich interface, where the \ce{Pb} atoms on the interface are coordinated to five \ce{I} and one \ce{Br} atoms (Figure ~\ref{fig:bulk_interfaces_definition}b).

We start by calculating the formation energies of these interfaces. First, the geometry of a \supercell{8}{2}{2} cubic supercell of \mixedsystem{0.5}{0.5} with one \ce{I}-rich interface and one \ce{Br}-rich interface is optimized using the same functional, PAW potentials, kinetic energy cutoff, and convergence criteria as in section~\ref{CsPbBr3_training} of the SI note~\ref{CsPbBr3}, and a \supercell{1}{4}{4} Monkhorst-Pack \textit{k}-grid. Using the volume of this optimized cell as the basis, structures with 2 interfaces of each kind are created, and positions of the atoms are optimized using the DFT parameters mentioned above. Finally, we calculate the formation energies of these interfaces ($E_{\mathrm{form}}$) as
\begin{equation}
E_{\mathrm{form}} = \frac{E_{\mathrm{int}} - [ xE_{\mathrm{CsPbI_{3}}} + (1-x)E_{\mathrm{CsPbBr_{3}}}]}{n_{int}}
\end{equation}
where $E_{\mathrm{int}}$ is the energy of the interface system, $x$ is the ratio of number of \ce{I} atoms to the total number of halide atoms in the system, $E_{\mathrm{CsPbI_{3}}}$ and $E_{\mathrm{CsPbBr_{3}}}$ are the energies of bulk \ce{CsPbI_{3}} and \ce{CsPbBr_{3}}, respectively, and $n_{int}$ is the number of unit interfaces in the system. The difference between the formation energies of both interfaces, i.e., $E_{\mathrm{form}}^{\text{Br-rich}}-E_{\mathrm{form}}^{\text{I-rich}}$ is very small, specifically \SI{2.5}{meV} or $0.1k_{\mathrm{B}}T$ at room temperature, therefore, we conclude that both of these interfaces are equally likely to form under equilibrium conditions.

Next, we perform five \SI{2}{ns} long MD runs at \SI{600}{K} using \supercell{16}{6}{6} cubic supercells of \mixedsystem{x}{1-x} with 2 interfaces of each kind and one halide point defect, such system sizes are chosen to minimize the interactions between the interfaces. The lattice constants are obtained using the procedure described in SI note~\ref{mixed_PT}. The structures are first equilibrated to the target temperature for \SI{100}{ps}, and both equilibration and production runs are performed in an \textit{NVT} ensemble using the same thermostat settings as described in SI note~\ref{mixed_migration}.

\clearpage

\section{Cubic domains}
To demonstrate the influence of defect occupancy at bulk interfaces on halide phase mixing, we perform \SI{10}{ns} long MD runs at \SI{600}{K} using \supercell{8}{8}{8} cubic supercells of \mixedsystem{x}{1-x} with \ce{I} cubic domain in the center of the \ce{Br} lattice and vice versa, with one halide point defect. These model structures consist of six \ce{I}-rich interfaces for \ce{I} cubic domain and six \ce{Br}-rich interfaces for \ce{Br} cubic domain. The lattice constants are obtained using the procedure described in SI note~\ref{mixed_PT}. The structures are first equilibrated to the target temperature for \SI{100}{ps}, and both equilibration and production runs are performed in an \textit{NVT} ensemble using the same thermostat settings as described in SI note~\ref{mixed_migration}. The final frames from these simulations are given in Figure 5 of the main text.

\clearpage

%%%%%%%%%%%%%%%%%%%%%%%%%%%%%%%%%%%%%%%%%%%%%%%%%%%%%%%%%%%%%%%%%%%%%
%% The appropriate \bibliography command should be placed here.
%% Notice that the class file automatically sets \bibliographystyle
%% and also names the section correctly.
%%%%%%%%%%%%%%%%%%%%%%%%%%%%%%%%%%%%%%%%%%%%%%%%%%%%%%%%%%%%%%%%%%%%%
\bibliography{SI}